\documentclass[draft=false,twocolumn,showpacs,preprintnumbers]{aa}
\usepackage{natbib,twoopt}
\usepackage{graphicx,units,multirow,subfigure,color,amsmath,amssymb,epsfig,soul}
\usepackage{courier}
\usepackage[varg]{txfonts}
\usepackage[usenames,dvipsnames,svgnames,table]{xcolor}
\usepackage[absolute,overlay]{textpos}

\newcommand{\beq}{\begin{equation}}
\newcommand{\eeq}{\end{equation}}
\newcommand{\ben}{\begin{eqnarray}}
\newcommand{\een}{\end{eqnarray}}
\newcommand{\besub}{\begin{subequations}}
\newcommand{\eesub}{\end{subequations}}
\newcommand{\bi}{\begin{itemize}}
\newcommand{\ei}{\end{itemize}}

\newcommand{\bea}{\begin{align}}
\newcommand{\eea}{\end{align}}

\newcommand{\citefig}[1]{Fig.~\ref{#1}}

\newcommand{\BIG}{{\sf BIG}}
\newcommand{\SLIM}{{\sf SLIM}}
\newcommand{\QUAINT}{{\sf QUAINT}}

\newcommand{\usine}{{\sc usine~v3.5}}
\newcommand{\minuit}{{\sc minuit}}
\newcommand{\minos}{{\sc minos}}
\newcommand{\usinebis}{{\sc usine}}

\newcommand{\het}{\ensuremath{^3}\textrm{He}}
\newcommand{\hef}{\ensuremath{^4}\textrm{He}}

\newcommand{\sevenBe}{\ensuremath{^7}\textrm{Be}}
\newcommand{\nineBe}{\ensuremath{^9}\textrm{Be}}
\newcommand{\tenBe}{\ensuremath{^{10}}\textrm{Be}}
\newcommand{\tentonineBe}{\tenBe{}/\nineBe{}}
\newcommand{\tentototBe}{\tenBe{}/\textrm{Be}}
\newcommand{\tenB}{\ensuremath{^{10}}\textrm{B}}

\newcommand{\chimindof}{\ensuremath{\chi^2_{\rm min}}/{\rm dof}}

\newcommand{\chipernui}{\ensuremath{\chi^2_{\rm nui}/{n_{\rm nui}}}}

\def\aap{A\&A}%
\def\apjl{ApJ}%
\def\apjs{ApJS}%
\def\ao{Appl.~Opt.}%

\bibpunct{(}{)}{;}{a}{}{,} 

\definecolor{light-gray}{gray}{0.95}
\definecolor{dark-gray}{gray}{0.4}

\def\pbar{\ensuremath{\overline{p}}}

\usepackage[pdftex,             
            breaklinks=true,%
            colorlinks=true,%
            citecolor=blue,
            pdfauthor={Weinrich et al.},%
            pdftitle={Halo size of the Galaxy}%
           ]{hyperref}

\makeatletter

\newcommandtwoopt{\citeads}[3][][]{\href{http://adsabs.harvard.edu/abs/#3}{\def\hyper@linkstart##1##2{}\let\hyper@linkend\@empty\citealp[#1][#2]{#3}}}
\newcommandtwoopt{\citepads}[3][][]{\href{http://adsabs.harvard.edu/abs/#3}{\def\hyper@linkstart##1##2{}\let\hyper@linkend\@empty\citep[#1][#2]{#3}}}
\newcommandtwoopt{\citetads}[3][][]{\href{http://adsabs.harvard.edu/abs/#3}{\def\hyper@linkstart##1##2{}\let\hyper@linkend\@empty\citet[#1][#2]{#3}}}
\newcommandtwoopt{\citealpads}[3][][]{\href{http://adsabs.harvard.edu/abs/#3}{\def\hyper@linkstart##1##2{}\let\hyper@linkend\@empty\citealp[#1][#2]{#3}}}
\newcommandtwoopt{\citealtads}[3][][]{\href{http://adsabs.harvard.edu/abs/#3}{\def\hyper@linkstart##1##2{}\let\hyper@linkend\@empty\citealt[#1][#2]{#3}}}
\newcommandtwoopt{\citeyearads}[3][][]{\href{http://adsabs.harvard.edu/abs/#3}{\def\hyper@linkstart##1##2{}\let\hyper@linkend\@empty\citeyear[#1][#2]{#3}}}
\newcommandtwoopt{\citeadsstar}[3][][]{\href{http://adsabs.harvard.edu/abs/#3}{\def\hyper@linkstart##1##2{}\let\hyper@linkend\@empty\citealp*[#1][#2]{#3}}}
\newcommandtwoopt{\citepadsstar}[3][][]{\href{http://adsabs.harvard.edu/abs/#3}{\def\hyper@linkstart##1##2{}\let\hyper@linkend\@empty\citep*[#1][#2]{#3}}}
\newcommandtwoopt{\citetadsstar}[3][][]{\href{http://adsabs.harvard.edu/abs/#3}{\def\hyper@linkstart##1##2{}\let\hyper@linkend\@empty\citet*[#1][#2]{#3}}}
\newcommandtwoopt{\citeyearadsstar}[3][][]{\href{http://adsabs.harvard.edu/abs/#3}{\def\hyper@linkstart##1##2{}\let\hyper@linkend\@empty\citeyear*[#1][#2]{#3}}}
\newcommandtwoopt{\citeauthoradsstar}[3][][]{\href{http://adsabs.harvard.edu/abs/#3}{\def\hyper@linkstart##1##2{}\let\hyper@linkend\@empty\citeauthor*[#1][#2]{#3}}}
\newcommandtwoopt{\citepthesis}[3][][]{\href{http://tel.archives-ouvertes.fr/docs/#3}{\def\hyper@linkstart##1##2{}\let\hyper@linkend\@empty\citep[#1][#2]{#3}}}
\newcommandtwoopt{\citetthesis}[3][][]{\href{http://tel.archives-ouvertes.fr/docs/#3}{\def\hyper@linkstart##1##2{}\let\hyper@linkend\@empty\citet[#1][#2]{#3}}}
\makeatother

\begin{document}

\begin{textblock*}{5cm}(15.8cm,2.8cm) 
   LAPTH-009/20
\end{textblock*}
\begin{textblock*}{5cm}(15.8cm,3.2cm) 
   LUPM:20-019
\end{textblock*}

\input epsf
\title{Galactic halo size in the light of recent AMS-02 data}

\author{N. Weinrich\inst{1}
  \and M. Boudaud\inst{2}\thanks{Deceased}
  \and L. Derome\inst{1}
  \and Y. G\'enolini\inst{3}
  \and J. Lavalle\inst{4}\thanks{\url{julien.lavalle@umontpellier.fr}}
  \and \\D. Maurin\inst{1}\thanks{\url{david.maurin@lpsc.in2p3.fr}}
  \and P. Salati\inst{5}\thanks{\url{pierre.salati@lapth.cnrs.fr}}
  \and P. Serpico\inst{5}
  \and G. Weymann-Despres\inst{1}
}

\institute{
LPSC, Universit\'e Grenoble Alpes, CNRS/IN2P3, 53 avenue des Martyrs, 38026 Grenoble, France
\and Instituto de F\'isica Te\'orica UAM/CSIC, Calle Nicol\'as Cabrera 13-15, Cantoblanco E-28049 Madrid, Spain
\and Niels Bohr International Academy \& Discovery Center, Niels Bohr Institute, University of Copenhagen, Blegdamsvej 17, DK-2100 Copenhagen, Denmark
\and LUPM, CNRS \& Universit\'e de Montpellier (UMR-5299), Place Eug\`ene Bataillon, F-34095 Montpellier Cedex 05, France
\and LAPTh, Universit\'e Savoie Mont Blanc \& CNRS, 74941 Annecy Cedex, France
}

\date{Received / Accepted}

\abstract
{The vertical diffusive halo size of the Galaxy, $L$, is a key parameter for dark matter indirect searches. It can be better determined thanks to recent AMS-02 data.}
{We set constraints on $L$ from Be/B and \tentototBe{} data, and we performed a consistency check with positron data. We detail the dependence of Be/B and \tentototBe{} on $L$ and forecast on which energy range better data would be helpful for future $L$ improvements.}
{We used \usine{} for the propagation of nuclei, and $e^+$ were calculated with the pinching method.}
{The current AMS-02 Be/B ($\sim3\%$ precision) and ACE-CRIS \tentototBe{} ($\sim 10\%$ precision) data bring similar and consistent constraints on $L$. The AMS-02 Be/B data alone constrain $L=5^{+3}_{-2}$~kpc at a 68\% confidence level (spanning different benchmark transport configurations), a range for which most models do not overproduce positrons. Future experiments need to deliver percent-level accuracy on \tentonineBe{} anywhere below 10 GV to further constrain $L$. }
{Forthcoming AMS-02, HELIX, and PAMELA \tentonineBe{} results will further test and possibly tighten the limits derived here. Elemental ratios involving radioactive species with different lifetimes (e.g. Al/Mg and Cl/Ar) are also awaited to provide complementary and robuster constraints.}


\maketitle

\section{Introduction}

Finding the Galactic cosmic-ray (CR) sources, solving the details of CR transport in the Galaxy, and using the CR as a channel to identify the nature of dark matter (DM) are among the main challenges in CR physics. In the last decade, anti-matter data received a lot of scrutiny, owing to the interpretation of the positron fraction rise as a DM signal (\citealt{2009Natur.458..607A}, \citealt{2008PhRvD..78j3520B}, and \citealt{2018ApJ...858..116F}), although standard astrophysics explanations are more likely (e.g. \citealt{HooperEtAl2009}, \citealt{2010A&A...524A..51D}, \citealt{2012APh....39....2S}, and \citealt{2019JCAP...04..024M}). Similarly, the presence of a DM contribution in the \pbar{} data is debated in the literature (\citealt{2018JCAP...01..055R}, \citealt{2019PhRvD..99j3026C}, \citealt{2019PhRvD..99j3014C}, and \citealt{Boudaud:2019efq}). The latter channel is actually one of the best at constraining weakly interacting massive particles DM candidates in the GeV-TeV mass range \citep[e.g.][]{2017NatPh..13..224C}.

DM interpretations for antimatter CRs depend on both the transport and geometry parameters (\citealt{2004PhRvD..69f3501D}, \citealt{2008PhRvD..77f3527D}, and \citealt{2016PhR...618....1A}); the latter is mostly determined via CR radioactive clocks. These clocks have a lifetime of approximately a million years, one order of magnitude shorter than the typical CR propagation time in the Galaxy. Ratios of a secondary (i.e. produced at the propagation stage only) unstable species to their stable counterpart, for instance \tentonineBe{}, allow one to break the degeneracy between the diffusion coefficient normalisation and the halo size of the Galaxy \citep[e.g.][]{2002A&A...381..539D}. However, CR isotopic separation in experiments is challenging. The first measurements were carried out more than forty years ago for $^{10}$Be \citep{1973Ap&SS..24...17W}, $^{36}$Cl \citep{1981ApJ...246.1014Y}, $^{26}$Al \citep{1982ApJ...252..386W}, and $^{54}$Mn \citep{1979ICRC....1..430W}; however, even recent measures (\citealt{1998ApJ...501L..59C}, \citealt{2001ApJ...563..768Y}, and \citealt{2004ApJ...611..892H}) are restricted to low energy mostly, that is, below a few hundreds of MeV/n. As an alternative and complementary approach, \citet{1998ApJ...506..335W} proposed to use elemental ratios (e.g. Be/B, Al/Mg) in which the CR clock appears both in the numerator (decayed fraction) and denominator (daughter fed by decaying CR). Elemental ratios have been measured up to hundreds of GeV/n, hence covering an energy range in which \tenBe{} goes from mostly decayed to meta-stable, with respect to the propagation time, at high energy. While waiting for the AMS-02 future release of Be isotope data, the high-precision Be/B ratio already gives useful constraints~\citep{2020PhRvD.101b3013E}. However, the isotopic fraction of \tenBe{} in B is a few percent only, and the sensitivity of Be/B to $L$ is partly drowned by the dominant presence of the stable nucleus \sevenBe{} \citep{2015PhRvC..92d5808T}.

The positron data have also been recently shown to provide interesting constraints on $L$ \citep{2014PhRvD..90h1301L}. In particular, for small $L$, the unavoidable secondary production, mostly from H and He CRs on the interstellar medium (ISM), may overshoot low-energy data points. Therefore they can be used as a complementary probe to set a lower limit on $L$.

This study follows up on our previous effort to determine CR transport parameters \citep{2019PhRvD..99l3028G,WeinrichEtAl2020}. We performed a joint analysis of Li/C, B/C, and various combinations of \tenBe{} data to determine $L$. We then drew models from the allowed regions of their parameter space and further checked their consistency with the positron constraint. We stress that in this analysis, as is the case in almost all similar studies, the halo size of the Galaxy is set to be a hard boundary where the CR density goes to zero.  This is an effective modelling of a more realistic picture that would probably involve a rapidly growing diffusion coefficient in the halo \citep{2013JCAP...03..036D,2015PhRvD..92h1301T}. In fact, a modelling of CR, gas, and wave interactions from first principles leads to the picture of a dynamic halo \citep{1991A&A...245...79B,1996A&A...311..113Z}. Recent results from \citet{2018PhRvL.121b1102E} show, in particular, that the turbulent cascade from CR sources and the self-generation of waves by CRs can introduce an effective halo size. Full numerical simulations aimed at accurately computing CRs in a magneto-hydrodynamic framework will probably give more insight into this problem, as envisaged in \citet{2020MNRAS.491..993G}. In the meantime, the concept of hard boundary remains a useful benchmark for many CR-related studies.

The paper is organised as follows: In Sect.~\ref{sec:setup}, we recall the propagation model and configurations used for the analysis. In Sect.~\ref{sec:constraintL_Be}, we assess the capability of \tentonineBe{}, Be/B, or B/C current data to determine $L$, accounting for various modelling uncertainties; we then provide the resulting constraints on $L$. In Sect.~\ref{sec:constraintL_other}, we discuss other possible observables to determine $L$, and we look at the flux of secondary positrons, given the Be/B constraint on $L$. In Sect.~\ref{sec:conclusions}, we summarise and conclude. Further details of the analysis and additional cross-checks are reported to the appendices: App.~\ref{app:rescaling} provides scaling relations of the transport parameters with $L$ for stable secondary species---they extend, for all transport parameters, those that were given in App.~C of \citet{2019PhRvD..99l3028G}; App.~\ref{app:Lper10Bedataset} highlights the constraints on $L$ set by various \tentonineBe{} datasets (low-energy experiments, ISOMAX, and preliminary PAMELA analysis); App.~\ref{app:7Beratio} outlines why $^7{\rm Be}/(^9{\rm Be}+^{10}\!{\rm Be})$, the most favourable isotopic ratio to extract experimentally, is of no practical use to constrain $L$.

\section{Model and configurations (\BIG{}, \SLIM{}, \QUAINT{})}
\label{sec:setup}

The details of the transport equation and approach we follow are detailed in \citet{2019PhRvD..99l3028G} and in the companion paper \citep{WeinrichEtAl2020}. Here, we only recall the most important features of the model.

We assume the CR density to obey a steady-state diffusion-advection equation. The geometry of the diffusion halo defines the region in which CRs propagate and are confined, here  considered an infinite slab of half-thickness $L$, the parameter we aim at determining in this analysis. In this geometry, CR sources and the gas are pinched in an infinitely thin plan (half-thickness $h=100$~pc$\ll L$), where interactions with the gas (spallation, energy gains and losses) are restricted to. A convection term $V_{\rm c}$ is taken to be constant and perpendicular to the disc. We assume isotropic and homogeneous diffusion, and account for diffusion in momentum space: we follow the treatment of \citet{1994ApJ...432..656S}, that is $K_{pp}(R,\vec{x})\propto 2\,h\,\delta(z)\,(V_{\rm A} p)^2/K(R)$ where $V_{\rm A}$ is the Alfv\'enic speed of the plasma wave. These assumptions allow one to derive solutions semi-analytically \citep{2001ApJ...547..264J,2001ApJ...555..585M}. Our calculations are performed with the code \usine{} \citep{2020CoPhC.24706942M}\footnote{\url{https://lpsc.in2p3.fr/usine}}. For more details, we refer the reader to \citet{2019PhRvD..99l3028G} and \citet{WeinrichEtAl2020}.

Several theoretical studies have hinted at the possible presence of breaks at low- \citep{2006ApJ...642..902P} or high-rigidity \citep{2012PhRvL.109f1101B,2018PhRvL.121b1102E}. Actually, spectral breaks are seen in CR data at low- \citep{2013Sci...341..150S} and high-rigidity \citep{2018PhRvL.120b1101A}, and they can be connected to the presence of breaks in the diffusion coefficient at rigidities $R_l\approx 4-5$~GV \citep{2019PhRvD..99l3028G,2019PhRvD.100d3007V,WeinrichEtAl2020} and $R_h\approx 250$~GV (\citealt{2017PhRvL.119x1101G}, \citealt{2018JCAP...01..055R}, \citealt{2019PhRvD..99l3028G}, and \citealt{2019PhRvD..99j3023E}). For these reasons we take
\begin{equation}
  \label{eq:def_K}
  K(R) = {\beta^\eta} K_{0} \;
  {\left\{ 1 \!+ \left( \frac{R}{R_{\rm l}} \right)^{\frac{\delta_{\rm l}-\delta}{s_{\rm l}}} \right\}^{s_{\rm l}}}
  {\left\{  \frac{R}{1\,{\rm GV}} \right\}^\delta}\,
  {\left\{  1 \!+ \left( \frac{R}{R_{\rm h}} \right)^{\frac{\delta-\delta_{\rm h}}{s_{\rm h}}}
    \right\}^{-s_{\rm h}}}\!\!\!\!\!\!.
\end{equation}
In this study, we fix several parameters whose impact on the results is negligible: the three high-rigidity break parameters ($R_h,\delta_h,s_h$) are set to the values reported in~\citet{2019PhRvD..99l3028G};  the smoothness of the low-rigidity break parameter is fixed at $s_l$ = 0.04 (fast transition).

\begin{table}
\begin{center}
\caption{Free ($\checkmark$) and fixed parameters for the benchmark configurations analysed in this study, see Eq.~(\ref{eq:def_K}). We schematically separate the parameters in several rigidity domains, but $V_{\rm A}$ and $V_{\rm c}$ impact fluxes across both the low- and intermediate-rigidity domain.}
\label{tab:free_params}
\begin{tabular}{cccc}
\hline\hline
Parameters   & \BIG{}    & \SLIM{}     & \QUAINT{}  \\
\hline
\multicolumn{4}{c}{\em Low-rigidity parameters}\\[5pt]
 $\eta$           & 1          &  1         & \checkmark \\
 $\delta_{\rm l}$ & \checkmark & \checkmark & n/a        \\
 $s_{\rm l}$      & 0.05       &  0.05      & n/a        \\
 $R_{\rm l}$      & \checkmark & \checkmark & n/a        \\
$V_{\rm A}$       & \checkmark & n/a        & \checkmark \\[5pt]
\multicolumn{4}{c}{\em Intermediate-rigidity parameters}\\[2pt]
$V_{\rm c}$       & \checkmark & n/a        & \checkmark \\
 $K_0$            & \checkmark & \checkmark & \checkmark \\
 $\delta$         & \checkmark & \checkmark & \checkmark \\[5pt]
\multicolumn{4}{c}{\em High-rigidity parameters}\\[2pt]
$\Delta_{\rm h}$  & $0.18$      & $0.19$    & $0.17$     \\
$R_{\rm h} $ [GV] & $247$       & $237$     & $270 $     \\
$s_{\rm h}$       & $0.04$      & $0.04$    & $ 0.04$    \\[5pt]
\multicolumn{4}{c}{\em Geometry parameter}\\[2pt]
$L$               & \checkmark & \checkmark & \checkmark \\
\hline
\end{tabular}
\end{center}
\end{table}
We use three benchmark configurations \BIG{}, \SLIM{}, and \QUAINT{} defined in \citet{2019PhRvD..99l3028G}, whose relevant parameters are collected in Table~\ref{tab:free_params}. In this analysis, \BIG{} has 7 free parameters $(K_0,\,\delta,\,R_l,\,\delta_l,\, V_{\rm c},\,V_{\rm A},\,L)$. The configuration \SLIM{} is a special case of \BIG{}, with $V_{\rm A}\!=\!V_{\rm c}\!=\!0$ and $\eta\!=\!1$, and it has 5 free parameters ($K_0,\,\delta,\,R_l,\,\delta_l,\,L)$. The configuration \QUAINT{} is also a special case of \BIG{} with no low-rigidity break, and it has 6 free parameters $(K_0,\,\delta,\,\eta,\, V_{\rm c},\, V_{\rm A},\,L)$. \QUAINT{} is also an extension (because of the high-rigidity break) of older benchmark convection and reacceleration models used for instance in \citet{2010A&A...516A..67M} and \citet{2010APh....34..274D}.

We fitted the model to the data via a $\chi^2$ minimisation,
\begin{equation}
  \chi^2 = \sum_{t}\left( \sum_{q_t} \left( {\cal D}_{\rm cov}^{t,q_t} + {\cal N}^{t,q_t}_{\rm Sol.Mod.}\right) \right) + \sum_r{\cal N}^{r}_{\rm XS},
  \label{eq:chi2}
\end{equation}
where $t$ and $q_t$ run over the several flux ratios (e.g. Li/C, Be/C, B/C) measured at different  periods, whereas $r$ runs over cross-section reactions. The non-diagonal ${\cal D}_{\rm cov}^{t,q_t}$ term allows for ${ij}$ energy bins correlations in the data (covariance matrix). The ${\cal N}^{t,q}_{\rm Sol.Mod.}$ and ${\cal N}_{\rm XS}$ terms account for Solar modulation and cross-section nuisance parameters, respectively. More details on this procedure are given in App.~B of \citet{2019A&A...627A.158D}. For further details related to the Li/C, Be/C, and B/C data uncertainties (correlation matrix of systematics), and also solar modulation and cross-section priors, we refer the reader to our companion paper \citep{WeinrichEtAl2020}.

The minimisation is performed with the \minuit\ package \citep{1975CoPhC..10..343J}, its \minos{} algorithm also providing accurate (asymmetric) error bars even if the problem is very non-linear. In practice, the \minuit{} routines are directly called from \usinebis{} \citep{2020CoPhC.24706942M}. We also carry out ${\cal O}(100)$ minimisations from different starting points to ensure the true minimum is found \citep{WeinrichEtAl2020}. All uncertainties reported on the halo size in this study are derived from the profile likelihood method (with \minos{}) at the 68\% confidence level. As the $\chi^2$ definition accounts for energy correlations in the data uncertainties and nuisance parameters, the halo size uncertainties also account for them. We stress that uncertainties were derived on $\log(L)$ and are mostly symmetric on this parameter. For this reason, we loosely use the notation $1\sigma$ in the following, and for instance, $2\sigma$ limits can be estimated assuming a log-normal distribution for $L$.

\section{Halo size $L$ from CR clocks}
\label{sec:constraintL_Be}

Radioactive secondary species whose lifetime is shorter than escape time decay before experiencing the boundary of the Galaxy. These species are only sensitive to the diffusion coefficient $K$, whereas stable secondary species can escape and are sensitive to $K/L$. Any fit combining the information of a stable and radioactive secondary species breaks the $K/L$ degeneracy, allowing for the determination of $L$ \citep{2002A&A...381..539D}.

Below, we focus on \tenBe{} ($t_{1/2}=1.387$~Myr) and related ratios (\tentonineBe{}, \tentototBe{}, and Be/B). To date, \tenBe{}  is the only available CR chronometer for which high-precision data exist for the associated elemental flux, Be. The Al, Cl, and Mn fluxes have not been released by the AMS-02 collaboration, yet.

\subsection{CR datasets and modulation levels}
\label{subsec:datasets}

This analysis is based on several datasets. Each dataset is associated to a solar modulation level depending on its data taking period. We use here the simple force-field approximation (\citealt{1967ApJ...149L.115G,1968ApJ...154.1011G}, and \citealt{2004JGRA..109.1101C}), whose single parameter,
$\phi_{\rm FF}$, is taken as a nuisance parameter in the analyses. As detailed in \citet{WeinrichEtAl2020}, the solar modulation level (prior) for each dataset is based on the analysis of neutron monitor data \citep{2015AdSpR..55..363M,2016A&A...591A..94G,2017A&A...605C...2G,2017AdSpR..60..833G}.

\begin{table}[t]
\caption{List of experiments with their data-taking periods and associated expected Solar modulation level. In the list below, datasets from the same experiment and data taking periods share a common Solar modulation level in the analyses. PAMELA data are from a preliminary analysis. For this reason, they are never considered in our main results, and are only used for illustration in App.~\ref{app:Lper10Bedataset}.}
{
\footnotesize
\label{tab:LiBeBC_data}
\begin{tabular}{rll}
\hline\hline
Experiment (period) & $\phi_{\rm prior}$      & Reference\\
                    &      [MV]               &          \\
\hline \\[-1em]
\multicolumn{3}{c}{\bf Li/C and B/C (`Base')}\\[2pt]
                  AMS-02 ('11/05-'16/05)  &  676           & \citet{2018PhRvL.120b1101A} \\
                ACE-CRIS ('97/08-'98/04)  &  528           & \citet{2013ApJ...770..117L} \\
                ACE-CRIS ('98/01-'99/01)  &  582           & \citet{2006AdSpR..38.1558D} \\
                ACE-CRIS ('01/05-'03/09)  &  872           & \citet{2013ApJ...770..117L} \\
                ACE-CRIS ('09/03-'10/01)  &  445           & \citet{2013ApJ...770..117L} \\
                    IMP8 ('74/01-'78/10)  &  540           & \citet{1987ApJS...64..269G} \\
               ISEE3-HKH ('78/08-'81/04)  &  742           & \citet{1988ApJ...328..940K}\!\!\!\!\!\!\!\! \\
\hspace{-9mm}Ulysses-HET ('90/10-'95/07)  &  732           & \citet{1996AaA...316..555D} \\
\hspace{-5mm}Voyager1\&2 ('77/01-'98/12)  &  450           & \citet{1999ICRC....3...41L} \\[6pt]
\multicolumn{3}{c}{\bf Be/B data}\\[2pt]
                  AMS-02 ('11/05-'16/05)  &  676           & \citet{2018PhRvL.120b1101A} \\[6pt]
\multicolumn{3}{c}{\bf \tentototBe{} data}\\[2pt]
                 IMP7\&8 ('72/09-'75/09)  &  543           & \citet{1977ApJ...217..859G} \\
                 IMP7\&8 ('74/01-'80/05)  &  580           & \citet{1981ICRC....2...72G} \\
               ISEE3-HKH ('78/08-'79/08)  &  653           & \citet{1980ApJ...239L.139W}\!\!\!\!\!\!\!\! \\
\hspace{-9mm}Ulysses-HET ('90/10-'97/12)  &  661           & \citet{1998ApJ...501L..59C} \\
\hspace{-5mm}Voyager1\&2 ('77/01-'98/12)  &  450           & \citet{1999ICRC....3...41L} \\[6pt]
\multicolumn{3}{c}{\bf \tentonineBe{} data}\\[2pt]
\multicolumn{3}{c}{\em Most precise low-energy data (`ACE')}\\[1pt]
                ACE-CRIS ('97/08-'99/07)  &  581           & \citet{2001ApJ...563..768Y} \\
                 ACE-SIS ('97/08-'99/07)  &  581           & \citet{2001ApJ...563..768Y} \\
\multicolumn{3}{c}{\em Remaining low-energy data (`LE w/o ACE')}\\[1pt]
                 IMP7\&8 ('74/01-'80/05)  &  580           & \citet{1981ICRC....2...72G} \\
               ISEE3-HKH ('78/08-'79/08)  &  653           & \citet{1980ApJ...239L.139W}\!\!\!\!\!\!\!\! \\
\hspace{-9mm}Ulysses-HET ('90/10-'97/12)  &  661           & \citet{1998ApJ...501L..59C} \\
\hspace{-5mm}Voyager1\&2 ('77/01-'98/12)  &  450           & \citet{1999ICRC....3...41L} \\
\multicolumn{3}{c}{\em Intermediate energy}\\[1pt]
                         ISOMAX ('98/08)  &  618           & \citet{2004ApJ...611..892H} \\
\multicolumn{3}{c}{\em Preliminary data (App.~\ref{app:Lper10Bedataset} only)}\\[1pt]
                   PAMELA ('06/07-'14/09) &  500           & \citet{2019BRASP..83..967B}\!\!\!\!\!\!\! \\
  \hline
\end{tabular}
}
\end{table}
As in \citet{WeinrichEtAl2020}, the baseline data used to fix the transport parameters are AMS-02 Li/C and B/C data \citep{2018PhRvL.120b1101A}---denoted `base' in the following. The halo size $L$ is then constrained by combining the base with several ratios from different datasets. The largest dataset is that of AMS-02 Be/B, covering $\sim3$~GV to $\sim 2$~TV.  We also have several low-energy \tentonineBe{} or \tentototBe{} datasets available, retrieved from CRDB\footnote{\url{https://lpsc.in2p3.fr/crdb}} \citep{2014A&A...569A..32M}. Except for the ISOMAX data reaching $\sim2$~GeV/n, most of them (ACE, IMP7\&8, ISEE3, Ulysses, and Voyager1\&2) are at a few hundreds of MeV/n.

The various datasets of interest are listed in Table~\ref{tab:LiBeBC_data}, along with their estimated Solar modulation level and bibliographic reference. Several ratio and dataset combinations are considered, in order to assess and compare their respective impact on $L$. Combined to `base' (i.e. Li/C and B/C data only), which enables the determination of the transport parameters, at least one \tenBe{}-related dataset is necessary to determine $L$. In particular, for fits involving \tentonineBe{}, we differentiate three groups (see Table~\ref{tab:LiBeBC_data}): the most precise low-energy data only (`ACE'), low-energy data without ACE (`LE w/o ACE'), or intermediate energy data only (`ISOMAX'). For consistency for the Solar modulation nuisance parameters, in all fits involving low-energy isotopic ratios, we consider, if available, the associated similarly modulated Li/C and B/C low-energy data (in addition to AMS-02 data).

\subsection{Expected constraints from B/C, Be/B, and $^{10}$Be/Be}
\label{subsec:expected}

Before moving to the fits and results, we wish to understand how strongly $L$ can be constrained by different data combinations of \tenBe{}. Ratios directly involving \tenBe{} should be optimal, but owing to the experimental difficulty of achieving isotopic separation, high-precision elemental ratios like Be/B can be competitive \citep{1998ApJ...506..335W}. In principle, B/C might also lead to some  constraint, via the fraction of \tenB{} generated by \tenBe{} decay. The most favourable option would include some information on the \tenBe{} content, high-precision data, and significant dynamic range. For instance, at $\sim 1$~GeV/n, the B/C and Be/C fraction from unstable isotopes are respectively $\sim5\%$ and $5-10\%$, with data available from GV to TV at $3-5\%$ precision \citep{2018PhRvL.120b1101A}; for \tenBe{}, the fraction is $100\%$ with data in the $\sim 50-200$~MeV/n range at $\sim 10-15\%$ precision \citep{2001ApJ...563..768Y}.

\begin{figure}[!t]
  \includegraphics[width=\columnwidth]{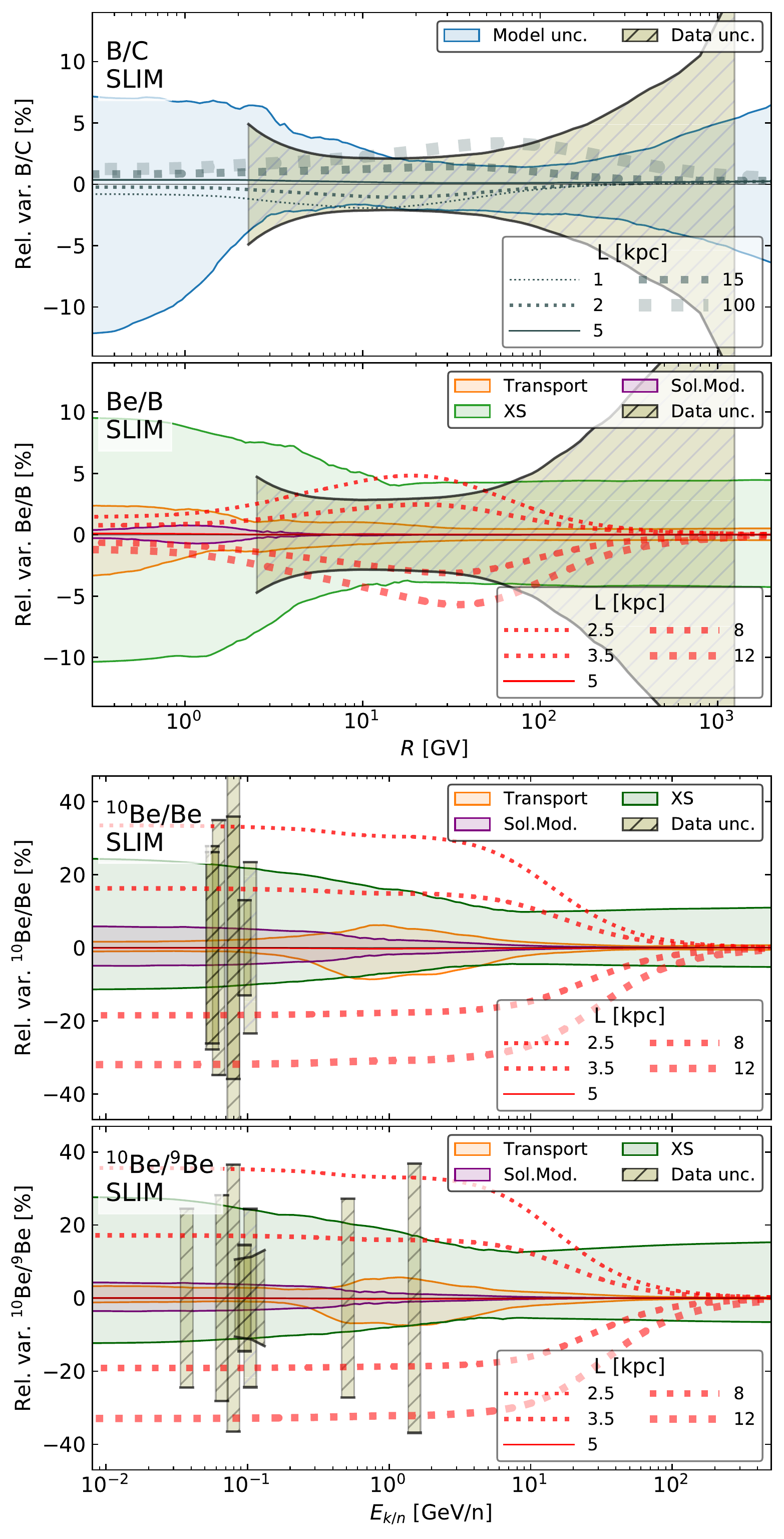}
  \caption{Relative variation of B/C, Be/B, \tentototBe{}, and \tentonineBe{} ratios (from top to bottom) as a function of rigidity (upper panels) or kinetic energy per nucleon (lower panels). All the panels compare three different effects: (i) the thin- to thick-dashed lines correspond, at constant grammage, to the impact of $L$ on the calculated ratios (the reference is $L=5$~kpc); (ii) hatched boxes show the data relative uncertainties over the data energy coverage (AMS-02 data in the two top panels) or for central energy points only (ACE, IMP7\&8, ISEE3, ISOMAX, Ulysses, and Voyager1\&2 data in the two bottom panels)---see Sect.~\ref{subsec:datasets} and Table~\ref{tab:LiBeBC_data} for data references; (iii) the remaining envelopes show $68\%$ CLs on model calculations, including transport, cross sections, and modulation uncertainties separately or combined (`total'). \label{fig:constraints_gilles_plot}
}
\end{figure}

We first pick a reference configuration and $L$ to which other calculations are compared. This configuration is taken from our companion paper, \citet{WeinrichEtAl2020}, and is based on the simultaneous analysis of AMS-02 Li/C, Be/C and B/C data at fixed $L=5$~kpc. To see how sensitive to $L$ calculated ratios are, we vary $L$ ensuring that the level of production remains unchanged for secondary stable species---this constraint is satisfied with an appropriate rescaling of the transport parameters (see App.~\ref{app:rescaling})\footnote{There is a subtlety for B/C, as it contains $^{10}$B having decayed from \tenBe{}. Depending how scaling relations are derived, they absorb or not the $L$ dependence of this decayed fraction. We use the theory-based scaling relations to study the sensitivity of B/C to $L$, but the fit-driven ones for the remaining ratios (see App.~\ref{app:rescaling}).}. From this set-up, we calculate the relative variation of B/C, Be/B, $^{10}$Be/Be, and $^{10}$Be/$^{9}$Be with respect to the reference values at $L=5$~kpc.

The results are shown in \citefig{fig:constraints_gilles_plot}, with the dependence upon growing $L$ illustrated via the growing thickness of dotted lines, while different uniformly coloured bands represent different theoretical uncertainties (described and discussed in Sec.~\ref{sec:future_efforts}). For B/C (top panel), even for unrealistic variations of $L$, the maximal impact is a few percent only. On the other hand, varying $L$ from 2.5 to 12~kpc leads to variations of up to $5\%$ for Be/B (second panel) and up to $35\%$ for \tentonineBe{} and \tentototBe{} (two bottom panels). Secondly, whereas these variations peak strongly around $\sim30$~GV for Be/B, they are constant for all energy below a few GeV/n for \tentonineBe{} and \tentototBe{}.
We overlay on the plots (hatched regions) the existing data coverage in terms of energy range and total uncertainties. From the comparison with the $L$-dependent curves (thin- to thick-dotted lines), we conclude that B/C data alone can only provide very loose upper limits ($\lesssim20-30$ kpc), while Be/B, \tentonineBe{}, and \tentototBe{} are all expected to constrain $L$ to better than a factor 2.

\subsection{Directions for future experimental efforts}
\label{sec:future_efforts}
By pursuing further the reasoning, we can also forecast where future measurements could improve the constraint on $L$.
From \citefig{fig:constraints_gilles_plot}, energy ranges where $L$ is the most impacting are between 10 and 100 GV for Be/B data, and below a few GeV/n for \tentonineBe{}. However, to fairly assess the sensitivity of each ratio, model uncertainties should be taken into account---broadly speaking, model uncertainties originate from transport, cross-section, and solar modulation.

Technically we proceed as follows: starting from the best fit and covariance matrix of the relevant parameters (standard outputs of fits with \usine{}), we draw realisations of the parameters of interest, from which new values for the ratios are calculated; we then extract contours and confidence levels on these ratios. Repeating the drawing procedure considering the full covariance matrix of parameters or only block elements of this matrix, we can propagate the model uncertainties all together (e.g. B/C in the top panel of \citefig{fig:constraints_gilles_plot}) or separately (all remaining panels). This naturally accounts for the full or partial correlations between the parameters.

We show in the panels of \citefig{fig:constraints_gilles_plot} various model uncertainties ($1\sigma$ contours) obtained from sampling $10^4$ realisations of the model parameters for \SLIM{}\footnote{Configurations having more free parameters (\QUAINT{} and \BIG{}) would provide larger uncertainties; see Fig.~4 in \citet{WeinrichEtAl2020}.}. Overall, solar modulation uncertainties ($\lesssim 5\%$, purple contours) are sub-dominant in the model error budget; they are also sub-dominant with respect to data uncertainties (hatched boxes). For isotopic ratios, cross-section uncertainties dominate ($\sim 10-20\%$, green contours), followed by transport uncertainties ($\lesssim 10\%$, orange contours). Let us detail separately the conclusions that can be drawn for B/C, Be/B, \tentototBe{}, and \tentonineBe{}---for completeness, we also show in App.~\ref{app:7Beratio} the $^7{\rm Be}/(^9{\rm Be}+^{10}{\rm Be})$ ratio, more easily measured but unfortunately not constraining for $L$.

Firstly, the status of the B/C ratio (top panel) is different from the other ones, because transport is calibrated on it. For this reason, the associated uncertainties are typically at the level of the data uncertainties. Given the poor sensitivity of B/C to $L$, at least a factor of ten improvement on data errors would be necessary to bring any improvement on $L$.

Secondly, for the Be/B ratio (second panel), the optimal region to constrain $L$ is from a few GV to a few tens of GV, exactly in the region where AMS-02 data have the smallest errors ($\sim 3\%)$. In this region, cross-section uncertainties (green contours) are at the $5\%$ level, meaning that better nuclear cross-section data could already slightly shrink the allowed $L$ values by $\sim 30\%$. Better Be/B data could strengthen the $L$ constraint, but only with cross-sections uncertainties brought to par. However, improving CR data systematics below the percent level is very challenging.

Thirdly, for the isotopic \tentototBe{} and \tentonineBe{} ratios (bottom two panels), we have to slightly change our estimation of cross-section uncertainties. Contrarily to elemental ratios, in which only the overall element production matters (sum of $^7$Be, \nineBe{}, and \tenBe{}), isotopic ratios directly depend on the associated isotopic production cross sections. To be conservative---though probably too pessimistic---, we draw each isotopic production cross sections (independently) within their expected uncertainty range\footnote{We follow the NSS prescription \citep{2019A&A...627A.158D,WeinrichEtAl2020}, where we vary the cross-section normalisation and low-energy behaviour (power-law slope) of the dominant production channel for each Be isotope. We use a dispersion $\sigma_{\rm Slope}=0.15$ for these reactions, and $\sigma_{\rm Norm}=0.25$, 0.20, and 0.15 for $^{12}{\rm C\!+\!H}\!\!\to^{10}$Be, $^{12}{\rm C\!+\!H}\!\!\to^{9}$Be, and $^{16}{\rm O\!+\!H}\!\to^{7}$Be respectively.}. As for Be/B, the cross-section uncertainties are the dominant modelling uncertainties. At variance with Be/B, any region below 10~GeV/n is equally suited to constrain $L$. In the high-energy end of this interval, current data uncertainties are larger than the model uncertainties, so that there is a small window for improvements on data to improve the constraints on $L$. Also, CR data on a large energy range should provide a better lever arm to handle cross-section uncertainties.

We can now conclude on the best way to improve the constraints on $L$ in the future. On the short term, forthcoming \tentonineBe{} AMS-02 and PAMELA data up to 10 GeV/n are the best candidates to improve the constraints on $L$. On a longer term, the easiest way to improving isotopic CR data would be to focus on GeV/n energies for this same ratio. Improving the precision of Be/B data would be only significant at the sub-percent level, but this is likely to remain difficult to achieve, even for future CR projects like HERD \citep{2019NPPP..306...85C}, ALADInO \citep{Aladino2019}, or AMS-100 \citep{2019NIMPA.94462561S}. In any case, be it for Be/B or isotopic ratios, the common limiting factor to all improvements are cross-section uncertainties.

\subsection{Actual constraints on $L$}\label{constraintsL}
In this section, we present the constraints on $L$ from various data combinations. The fitting procedure and free (and nuisance) parameters are as discussed in Sect.~\ref{sec:setup}---see also our companion paper for more details \citep{WeinrichEtAl2020}.

\subsubsection{Results}

\begin{figure}[!t]
  \includegraphics[width=\columnwidth]{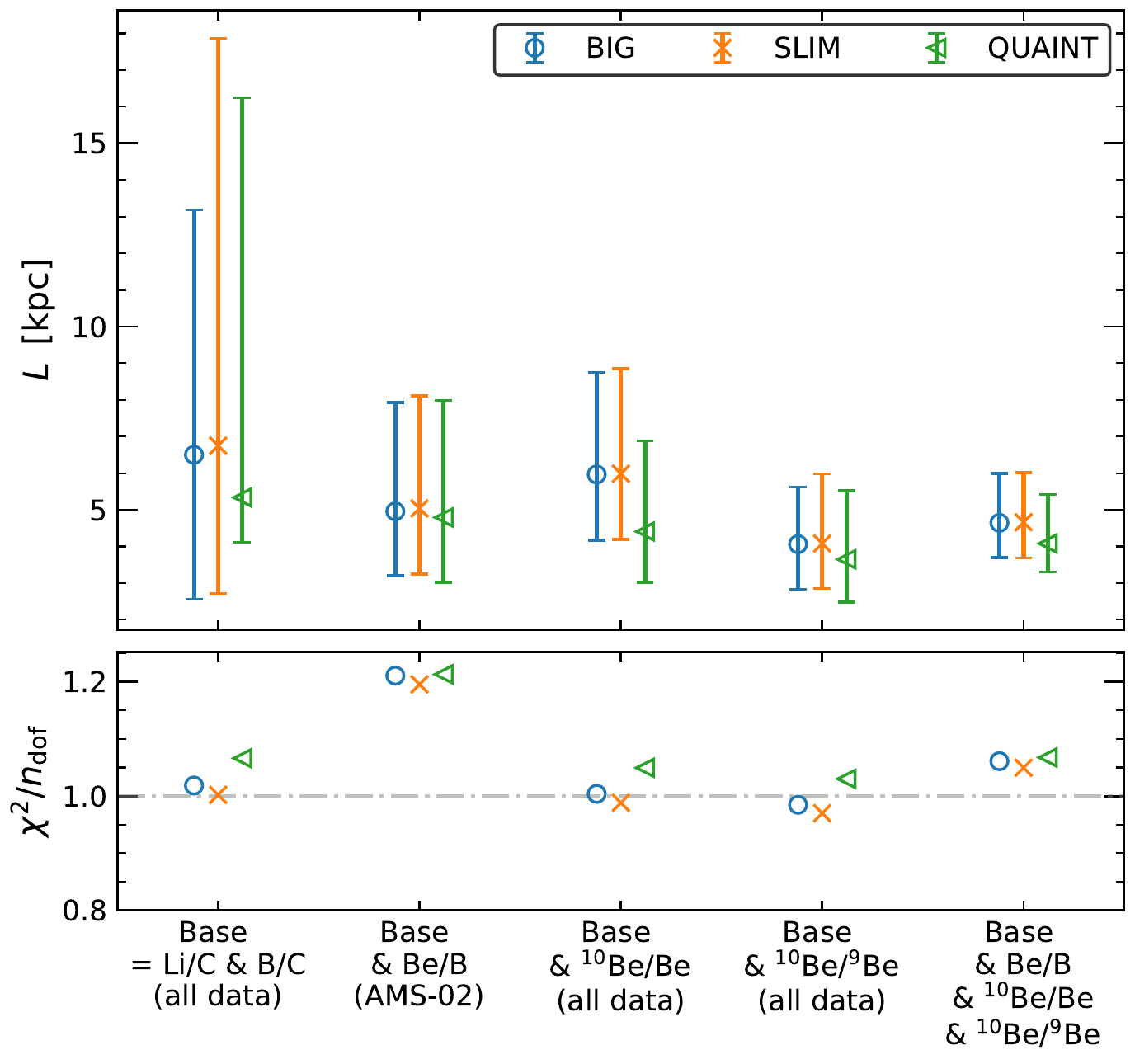}
  \caption{Best-fit halo size and asymmetric uncertainties (from \minos{}) for the configurations \BIG{}, \SLIM{}, and \QUAINT{}. From left to right, several data combinations are used. The first column (`Base') involves AMS-02 and low-energy B/C and Li/C data. The second column is the constraint set from AMS-02 data only (Li/C, Be/B, and B/C). The remaining columns combine `Base' data (from first column) to \tentototBe{} low-energy data (third column), to \tentonineBe{} low- and intermediate-energy (next-to-last column), or combine all the previous cases (last column).}
  \label{fig:FitVal_L}
\end{figure}
Figure~\ref{fig:FitVal_L} shows the constraints (at $1\sigma$) brought by the ratios discussed in the previous section (B/C, Be/B, \tentototBe{}, and \tentonineBe{}) for the three configurations \BIG{} (blue circle), \SLIM{} (orange crosses), or \QUAINT{} (green triangles). We first stress that the various transport configurations all give similar constraints on $L$---we do not show results for the transport parameters as they are available and were abundantly discussed in \citet{WeinrichEtAl2020}. The behaviour for the different ratios (columns) is in qualitative agreement with the expectations discussed in Sect.~\ref{subsec:expected}. In the first column, the B/C ratio mostly gives an upper limit $L\lesssim 15$~kpc, whereas in the second and third columns, Be/B AMS-02 and \tentototBe{} low-energy data give similar constraints $L\approx5\pm3$~kpc. The fourth column is based on the results from the same experiments as in the third column, but fitting \tentototBe{} instead of \tentonineBe{} data, and with the additional use of intermediate-energy ISOMAX data (see Table~\ref{tab:LiBeBC_data}). This gives a slightly lower best-fit value and uncertainties compared to \tentototBe{} data, but this apparent improvement is related to a tension between ISOMAX data and all the others. The reduced uncertainties result from an attempt to accommodate all the data at once---App.~\ref{app:Lper10Bedataset} details results on the broken-down constraints from various low-energy datasets. The tension with ISOMAX data also reflects in the global fit (last column), which is pushed towards slightly larger $L$ values, also preferred by AMS-02 Be/B data (second column).

\begin{table}[t]
  \centering
  \caption{Halo size fit results for the combined analysis of Li/C and B/C (denoted `Base', see also Fig.~\ref{fig:FitVal_L}) with an `unstable-to-stable' secondary ratio $r$. The top rows show the constraint from AMS-02 data ($r={\rm Be/B}$), while the bottom rows show the combined constraint from all available datasets ($r=$~Be/B~+\tentototBe{}~+\tentonineBe{}).}
  \label{tab:Geo_Lr}
\begin{tabular}{r ccc}
	\hline \hline \\[-1em]
	& \BIG{} & \SLIM{} & \QUAINT{}\\ \hline  \\[-1em]
  \multicolumn{4}{c}{\bf  Base \& Be/B}\\
  \multicolumn{4}{c}{\bf (AMS-02)}\\[3pt]
  $L$ [kpc]
  &$ 4.96^{+2.97}_{-1.76} $
  &$ 5.04^{+3.07}_{-1.79} $
  &$ 4.79^{+3.19}_{-1.77} $
  \\[3pt]
  $\chi^2 \,/\, n_\mathrm{dof}$
  & 233.7$\,/\,$193
  & 233.1$\,/\,$195
  & 235.3$\,/\,$194
  \\[2pt] 
  $\chi^2_\mathrm{nui} \,/\, n_\mathrm{nui}$
  & 17.4$\,/\,$20
  & 17.4$\,/\,$20
  & 15.8$\,/\,$20
  \\[10pt] 
	\multicolumn{4}{c}{\bf Base \& Be/B \&  \tentototBe{} \& \tentonineBe{}}\\
  \multicolumn{4}{c}{\bf (all data)}\\[3pt]
	$L$ [kpc]
	&$ 4.64^{+1.35}_{-0.94} $
	&$ 4.66^{+1.35}_{-0.97} $
	&$ 4.08^{+1.33}_{-0.78} $
	\\[3pt]
	$\chi^2 \,/\, n_\mathrm{dof}$
	& 266.3$\,/\,$251
	& 265.6$\,/\,$253
	& 269.0$\,/\,$252
	\\[2pt] 
	$\chi^2_\mathrm{nui} \,/\, n_\mathrm{nui}$
	& 25.6$\,/\,$35
	& 25.4$\,/\,$35
	& 25.6$\,/\,$35
	\\[2pt]
	\hline\\[-1em] 
\end{tabular}
\end{table}

We gather in Table~\ref{tab:Geo_Lr} the best-fit values and $1\sigma$ uncertainties on $L$ for the AMS-only analysis (with Be/B, top) and the combined analysis (with Be/B and all isotopic ratios, bottom). In terms of the \chimindof{} values, a fair but not perfect agreement is obtained when using AMS-02 only data ($\chimindof{}\sim1.2$).
An excellent fit is obtained for the isotopic data with $\chimindof{}\sim1.0$, and also when combining elemental and isotopic data with $\chimindof{}\sim1.06$ (last column in \citefig{fig:FitVal_L} or bottom of Table~\ref{tab:LiBeBC_data}); for the latter, low-energy Li/C, B/C, and also \tenBe{}-related ratios are in good agreement with the constraints set by AMS-02 data only and thus merely increases $n_{\rm data}$ without increasing $\chi^2_{\rm min}$.
The last row in Table~\ref{tab:Geo_Lr} shows the value of
\begin{equation}
  \chipernui{} \equiv \left(\sum_{s=0}^{n_s}{\cal N}^s_{\rm Sol.Mod.} + \sum_{x=0}^{n_x}{\cal N}^{x}_{\rm XS}\right)/(n_s+n_x),
  \label{eq:chi2nuis}
\end{equation}
with ${\cal N}^s_{\rm Sol.Mod.}$ and ${\cal N}^{x}_{\rm XS}$ the $n_s$ and $n_x$ nuisance parameters for solar modulation and cross sections respectively ($n_{\rm nui}=n_s+n_x$). As discussed in \citet{WeinrichEtAl2020}, this quantity gives a direct check that nuisance parameters behave properly. On average, nuisance parameters post-fit values should never be more than $1\sigma$ away from their prior, that is, $\chipernui{}\lesssim 1$, and this is verified for all our fits.

\begin{figure}[!t]
  \includegraphics[width=\columnwidth]{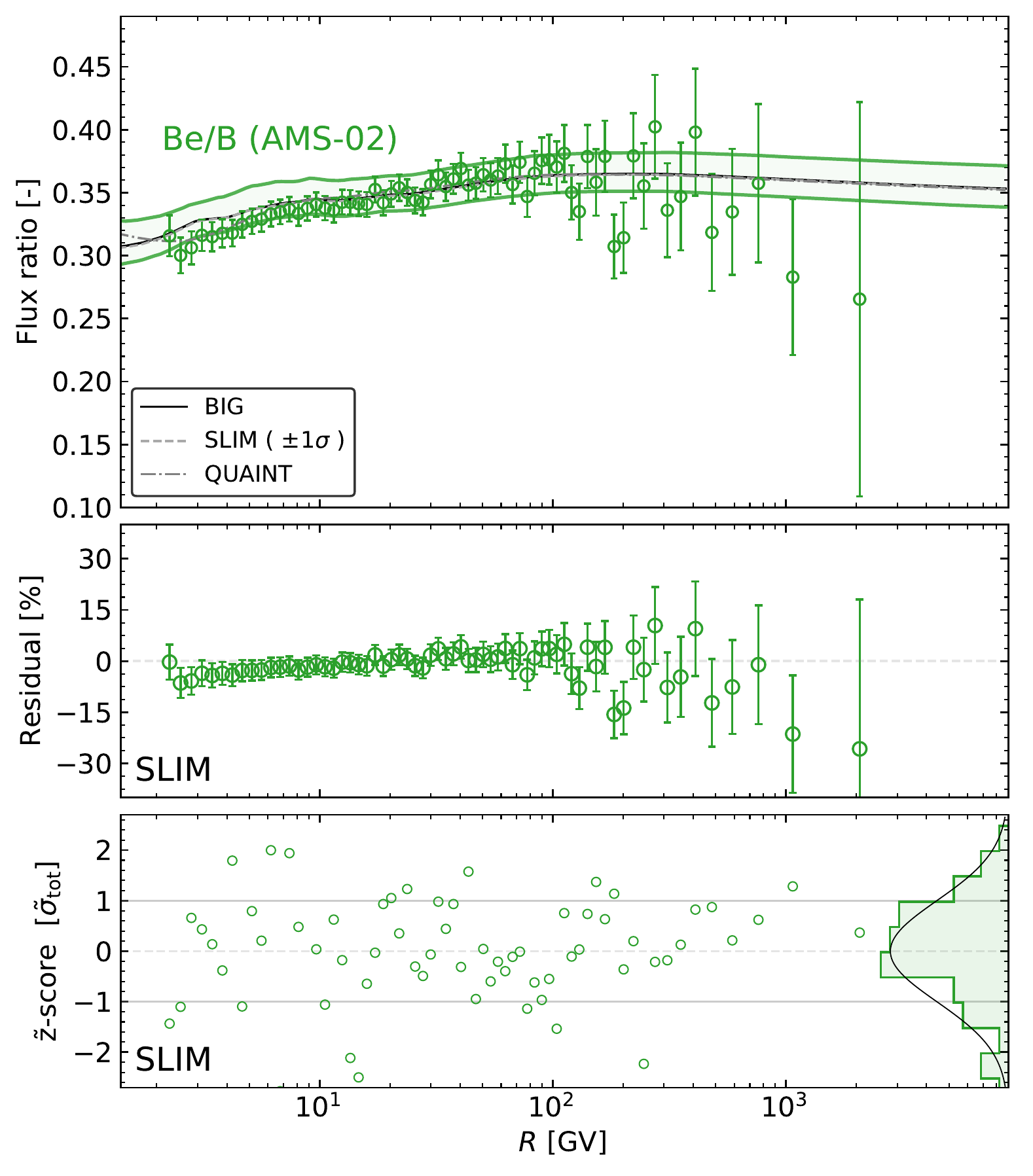}
  \caption{Model prediction (top), residuals (centre), and $\tilde{z}$-score (bottom) for Be/B based on the best-fit parameters to B/C, Li/C, \tentonineBe, \tentototBe{} and Be/B data. In the top panel, the contours show the $1\sigma$ total model uncertainties for \BIG{}. In the bottom panel, the right-hand side shows the distribution of $\tilde{z}$ values against a Gaussian with unit width (solid lines).}
  \label{fig:Be-contours}
\end{figure}
\begin{figure}[!t]
  \includegraphics[width=\columnwidth]{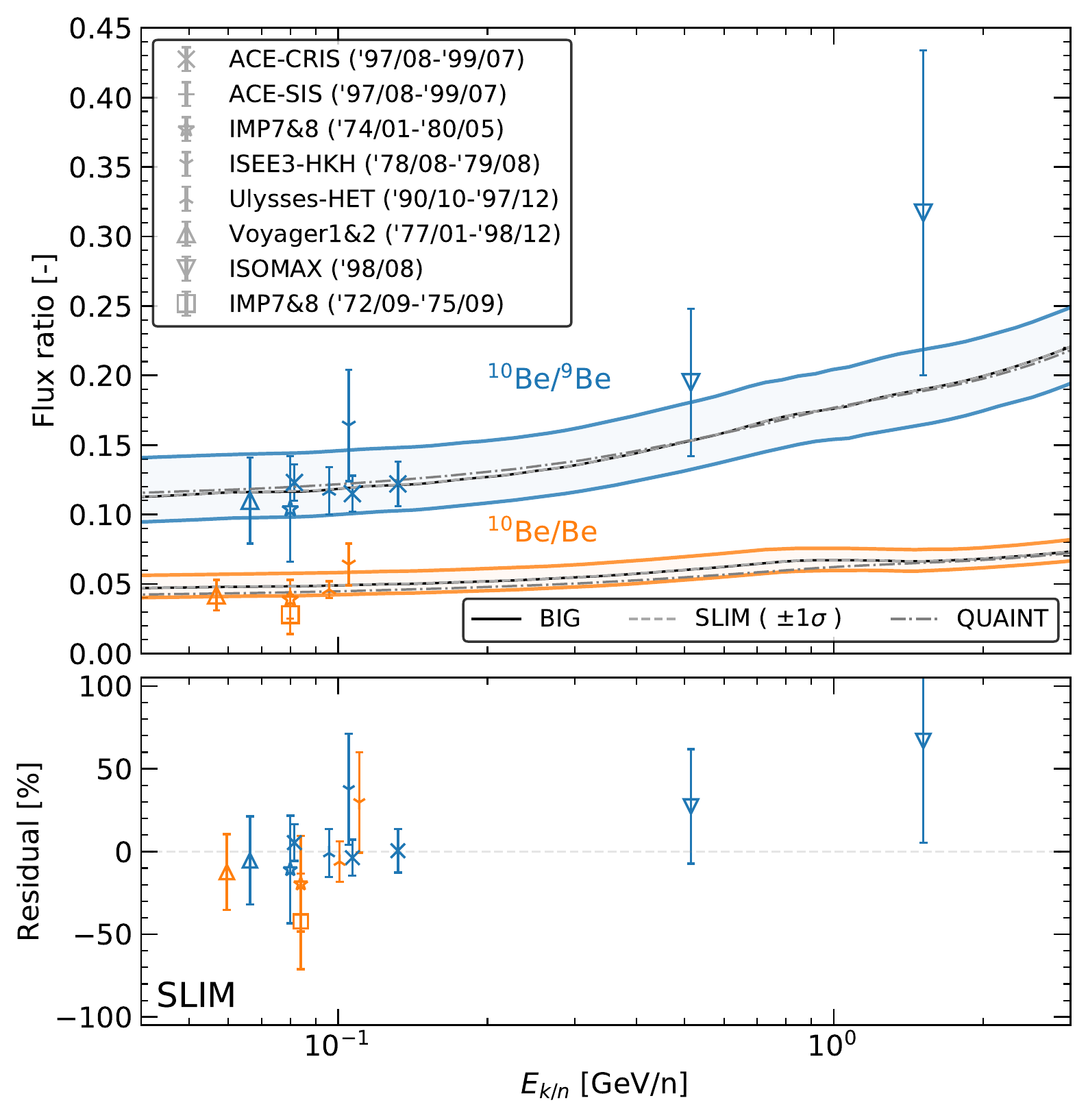}
  \caption{Model prediction (top) and residuals (bottom) for \tentonineBe{} (blue curves and symbols) and \tentototBe{} (orange curves and symbols). In the top panel, the $1\sigma$ contour corresponds to the total model uncertainties for \SLIM{} (as calculated in Sect.~\ref{sec:future_efforts}). The model calculations are based on the parameters from the combined fit of all Li/C, Be/C, Be/B, \tentonineBe{}, and \tentototBe{} data presented in Table~\ref{tab:LiBeBC_data} (PAMELA preliminary data excepted). }
  \label{fig:Beisot-contours}
\end{figure}
For illustration purposes, we finally show in Figs.~\ref{fig:Be-contours} and ~\ref{fig:Beisot-contours} the model calculation and the data for Be/B and isotopic ratios. The parameters are taken from the best-fit to all combined Be/B, \tentototBe{}, and \tentonineBe{} data (last column in \citefig{fig:FitVal_L}). In both plots, the top panels show the model calculations for the three transport configurations (\BIG, \SLIM, and \QUAINT) along with the data. For \SLIM, we also superimpose the $1\sigma$ model total uncertainties (contours) as calculated in Sect.~\ref{sec:future_efforts}. The second panels illustrate the goodness-of-fit to the data via the residuals between the data and the model. For the Be/B case with AMS-02 data (\citefig{fig:Be-contours}), a third panel shows the `rotated' score $\tilde{z}$, as defined in ~\citet{Boudaud:2019efq} or \citet{WeinrichEtAl2020}: It suffices to say that this score represents an unbiased visual representation of the distance between the model and the data, accounting for existing rigidity correlations in the systematics of AMS-02 data; also, the chi-square is the sum of the squares of these rotated residuals, that is, $\chi^2_{\rm Be/B}=\sum_i \tilde{z}_i^2 $. The right-hand side of the bottom panel is another illustration of the goodness of fit of the model, for the distribution of $\tilde{z}$-values is expected to follow a Gaussian distribution of width one.

\subsubsection{Discussion}

It is interesting to compare our results to those of previous analyses that considered either ACE-CRIS \tentonineBe{} ratio or Be/B data. 
Based on the analysis of \tentonineBe{} and other radioactive isotopes, and using a diffusion model with $\delta\approx0.3$, the GALPROP team found $L\in [1.5-6]$~kpc \citep{Moskalenko2001}, and later on, using an evolved Bayesian analysis, found $L=5.4 \pm 1.4$~kpc \citep{2011ApJ...729..106T}. Actually, the halo size strongly correlates with the diffusion slope $\delta$  \citep{2002A&A...381..539D,2010A&A...516A..66P}. Using 1D or 2D semi-analytical models, our team found $L\approx 5$~kpc \citep{2002A&A...381..539D} for $\delta\sim0.5$, and later on, also in an evolved Bayesian context, found $L\approx4\pm 1$~kpc in a pure diffusion/reacceleration model \citep{2010A&A...516A..66P}. All these values are consistent with the constraints derived here using ACE-CRIS \tentonineBe{} data only (see Fig.~\ref{fig:FitVal_L}), that is $L\in [3-8]$~kpc. Our uncertainties are larger than in previous studies, because we include here production cross-section uncertainties.

Comparatively, less studies focused on elemental ratios. Using HEAO-3 Be/B data (and other ratios) in a semi-analytical diffusion model with $\delta=0.6$, a rough range of $L\in [2-4]$~kpc was found in \citet{1998ApJ...506..335W}. A much larger range was found in \citet{2010A&A...516A..66P}, with $L$ a few kpc only allowed for $\delta\lesssim0.3$, growing to larger than 10~kpc for $\delta\approx 0.5$. This tension with \tentonineBe{} data was attributed to cross-section uncertainties---their importance in limiting the prediction power of Be/B is detailed in \citet{2015PhRvC..92d5808T}. Using updated cross sections and propagating their uncertainties, we find here $L=5^{+3}_{-2}$~kpc based on AMS-02 Be/B data, which is now compatible with values derived from ACE-CRIS \tentonineBe{} data. Similar values were found by the GALPROP team, using both ACE-CRIS and AMS-02 Be fluxes (without \tentonineBe{}), with $L=4\pm0.6$ \citep{2020ApJ...889..167B}. This stronger constraint is somehow in line with the fact that combining ACE-CRIS and AMS-02 data lead to tighter constraints (see Table~\ref{tab:Geo_Lr}), but as these authors do not use \tentonineBe{}, it is most probably attributed to a larger extent to the fact that they do not propagate cross-section uncertainties. In any case, the value of $L$ with a small error inferred in \citet{2020ApJ...889..167B} is also in mild tension with the recent analysis of AMS-02 Be data in \citet{2020PhRvD.101b3013E}, where a best-fit value of $L=7$~kpc but a lower limit $L\gtrsim 5$~kpc is found in a semi-analytical model similar to the one used here. Their central value, using total or statistical uncertainties only, is always enclosed within our $1\sigma$ confidence interval: firstly, the difference with our best-fit value is possibly due to their cross-section parametrisation \citep{2019PhRvD..99j3023E}, as illustrated in Fig.~2 of \citet{2020PhRvD.101b3013E}; secondly, our broader confidence interval is certainly related to our treatment (full propagation) of cross-section uncertainties.

In conclusion, most past and present analyses of \tentonineBe{} and Be/B show a preference for $L\approx4-5$~kpc, with a currently estimated $1\sigma$ uncertainty of about 50\%. Several analyses have combined or compared these results to the constraints brought by other ratios of radioactive secondary species (\citealt{1998ApJ...506..335W}, \citealt{Moskalenko2001}, \citealt{2002A&A...381..539D}, and \citealt{2010A&A...516A..66P}): similar halo sizes were observed, though with a large scatter. Forthcoming AMS-02 data on other elemental ratios (e.g. Al/Mg, Cl/Ar\dots) will allow one to repeat these analyses. Another effect could impact these conclusions. In \citet{2002A&A...381..539D} and \citet{2010A&A...516A..66P}, the impact of a local sub-density, exponentially attenuating the flux of radioactive species, was inspected: slightly larger or much larger uncertainties were observed depending on the transport configuration used \citep{2010A&A...516A..66P}. However, this was calculated assuming a similar diffusion in the local bubble and in the rest of the disc and halo, which may be questionable. Recent $\gamma$-ray observations \citep{2017Sci...358..911A,2019PhRvD.100l3015D}---interpreted as pockets of slow diffusion around pulsars \citep[e.g.][]{2018PhRvD..97l3008P}---, and the indication that the local ISM properties are affected by several SN explosions a few Myr ago \citep{2019BAAS...51c.410F} provide enough motivation to revisit this issue in a future study.

\section{Constraints on $L$ from other probes}
\label{sec:constraintL_other}

In this section, we discuss several independent constraints on $L$. We first review a series of constraints derived from multi-wavelength observations of the Milky Way and other spiral galaxies, and then determine explicitly direct constraints induced by low-energy secondary CR positrons.

\subsection{Direct and indirect constraints from radio and gamma's}
Radio emission in the MHz to GHz band from our Galaxy has been used to constrain the magnetised halo thickness $L$~(\citealt{2012JCAP...01..049B}, \citealt{2013MNRAS.436.2127O}, \citealt{2013JCAP...03..036D}, and \citealt{BiswasEtAl2018}). This band is dominated by the synchroton emission of the leptonic component of CR  in the Galactic magnetic field. The Galactic latitude profile of the radio maps are sensitive to the vertical gradient of CR sourcing the emission, $\propto 1/L$. Although relying on some assumptions (mostly on the magnetic field) these arguments tend to agree in excluding  low values of $L$, typically obtaining $L\gtrsim 2\,$kpc, with variations within a factor 2 depending on the analysis. They are somewhat less sensitive to large values of $L$, although upper limits in the range $L\lesssim 10-15\,$kpc have been derived. Further arguments based on radio observations such as rotation measurements of pulsars also yield results broadly consistent with these constraints~\citep{2013JCAP...03..036D}.

A complementary indirect indication can be derived by looking at the radio emissions of other spiral galaxies seen almost edge-on, which present an average scale-height of their synchrotron emission of about $1.8\pm0.2\,$kpc~\citep{2014arXiv1401.1317K}. This  translates, under the hypothesis of energy equipartition between magnetic field and cosmic ray energy density, in the typical constraint on the magnetised halo size $L\gtrsim 6.2-7.8\,$kpc~\citep{2015A&ARv..24....4B}. A recent analysis of an in-depth view of a spiral galaxy very similar to the Milky Way, NGC 891, reaches similar conclusions as for the vertical extension of the magnetic halo \citep{2019A&A...632A..12S}.

In principle, independent constraints can be derived from diffuse Galactic $\gamma$-ray data \citep{1977ApJ...217..843S}. The advantage is that the bulk of the data in the Fermi-LAT energy range (GeV) comes from $\pi^0$ decays of hadronic origin. Hence, the $\gamma$-ray flux mostly probes the convolution of the hadronic CR flux with the gas density. For instance, let us consider a simple model where CR transport is purely diffusive within a slab (1D-model). Assuming the gas is exponentially distributed with typical thickness h, the photon flux originating from the Galactic zenith scales as:
\begin{align}
\Phi_\gamma&\propto \int_0^\infty{\rm d}z\;\Phi_p(z)\;n_{\rm gas}(z)\nonumber\\&\propto \Phi_p(z=0)\int_0^L {\rm d}z \left(1-\frac{z}{L}\right)\; \exp\left(-\frac{z}{h}\right)\nonumber\\&\propto 1-\frac{h}{L}+{\cal O}\left(\frac{h^2}{L^2}\right)\,.
\end{align}
Since we expect $h/L\sim1\%-10\%$, the factors entering the actual $\gamma$-ray flux normalisation should be known to an unrealistically high precision for this method to provide meaningful constraints on $L$. However, additional inverse Compton contributions from leptonic CR component are also sensitive to $L$, and cannot be neglected. A global analysis of the diffuse $\gamma$-ray emission measured by the Fermi-LAT satellite actually shows a systematic improvement of the statistical likelihood as $L$ increases up to $\sim 10$ kpc \citep{2012ApJ...750....3A}, an effect particularly important when fitting the emission at large longitudes. Conversely, the study of high- and intermediate-velocity clouds at few kpcs away from the Galactic plane leads to the constraint $L< 6\,$kpc \citep{2015ApJ...807..161T}. Although this bound can be subject to variations (e.g. from the presence of unaccounted ionised gas in these clouds), the different trends between this constraint and the one obtained from radio observations of distant galaxies might be explained by the radio emission of leptons leaking out beyond the confinement volume. Also, the too shallow $\gamma$-ray gradient suggested by Fermi-LAT data admits alternative explanations, for instance in terms of a physically motivated correlation of the diffusion properties with galactocentric distance~\citep{2012PhRvL.108u1102E}.

\subsection{Constraints from $e^+$}
\label{sec:L_posit}

Secondary positrons, usually believed to dominate the local CR positron flux at low energy, can also be used to derive lower limits on the halo size (\citealt{2014PhRvD..90h1301L}, \citealt{BoudaudEtAl2017a}, and \citealt{2018JCAP...01..055R}). Their predicted abundance depends much less on the configuration of the Galactic magnetic field than the radio limits, and their theoretical uncertainties are better under control (see below). Secondary positrons may therefore significantly constrain propagation models, while still on the conservative side since their local flux is known to be dominated by primaries above a few GeV (e.g. \citealt{AharonianEtAl1995}, \citealt{2009Natur.458..607A}, \citealt{HooperEtAl2009}, and \citealt{2010A&A...524A..51D}).

\subsubsection{Positron flux scaling with $K_0$ and $L$ and calculation}
The steady-state local positron density is expected to scale like the production rate times the minimal propagation timescale involved. An additional dilution factor comes from the fact that the production volume $V_{\rm p}$ is, in most cases, smaller than the diffusion volume $V_\lambda$ \citep{BulanovEtAl1976}. Sticking to a one-dimensional picture, we have $V_{\rm p}/V_\lambda\approx h/\lambda(E)$. In the latter expression, $h$ is the half-height of the thin disc (where the ISM gas is confined), and $\lambda(E)= \sqrt{2\,K_{0}\,\tilde \tau(E)}$ is the positron propagation length scale, featuring the pseudo-energy loss timescale $\tilde \tau$.

The interesting regime to constrain $L$ is, similarly to radioactive species, when $h<\lambda<L$, that is, when the vertical boundary does not affect the positron density and $\tilde \tau$ does not depend appreciably on $K_0$. This typically happens at energies $\lesssim 10$~GeV, for which $\lambda \propto \sqrt{K_0}$, and the $e^+$ flux then scales as $1/\sqrt{K_0}$.
Therefore, since energy loss parameters are fixed independently from the propagation model, the positron flux is a direct probe of the diffusion coefficient normalisation $K_0$. Since the B/C ratio provides constraints on the ratio $K_0/L$, positron measurements enable an indirect probe of $L$: The lower $L$ in a B/C-compatible model, the larger the secondary positron flux.

In practice, we calculate the positron flux according to the pinching method introduced in \citet{BoudaudEtAl2017a}---see also \citet{DelahayeEtAl2009} for earlier attempts. For production, we consider incident and target species up to He only, and we take the cross-section parametrisation of \citet{KamaeEtAl2006}, which accounts for the low-energy hadronic resonances. Positron fluxes are then compared to AMS-02 data \citep{2019PhRvL.122d1102A} assuming $\phi_{\rm FF}=650$~MV, as estimated for the corresponding data-taking period May~2011-November~2017 (see below for a discussion on Solar modulation level and its uncertainties).
There are various sources of uncertainties in the calculation, and we try to list and quantify them below.

\subsubsection{Error budget}
We consider the uncertainties on the interstellar (IS) flux, and then comment on how the conclusions change for Top-of-Atmosphere (TOA) fluxes. We mostly focus on results at 1~GeV, because this is the typical energy where our analysis can draw constraints (see next subsection).

Firstly, we consider model uncertainties. Indeed, the method of calculation itself has some limitations. In the low- and high-energy regimes, propagation is dominated by energy losses in the disc and in the halo respectively. The pinching method allows to calculate intermediate energies by pinching the halo losses in the disc \citep{BoudaudEtAl2017a}, ensuring that both the limiting cases are recovered. Further comparisons against full numerical solutions should be carried out to definitively assess the accuracy of the method in the transition zone. Nevertheless, we have checked that the method is robust in the energy range used to define our limits below.

In a broader context, one could question the reliability of 1D models for consistency checks between nuclei and leptons, in the context of spatially-dependent distributions of sources and gas. Given the timescales of various transport parameters and energy losses, both these species originate from a few kpc away at GeV energies (\citealt{2003A&A...402..971T}, \citealt{2003A&A...404..949M}, and \citealt{2016ApJ...824...16J}). It means that their production and losses are sensitive to kpc-averages over the gas density properties---of course, this is no longer the case for very high energy leptons. So as long as we focus on the multi-GeV energy range, we do not expect strong differences due to gas inhomogeneities between nuclei and positrons, but as would be expected from more refined models, 1D model calculations are sensitive to the absolute value of the averaged gas density. This is at variance with the case of secondary radioactive species, discussed in ~Sec.~\ref{constraintsL}, which could be very sensitive to the local ISM. However, similarly to radioactive nuclei, inhomogeneous spatial diffusion zones around CR sources \citep{2017Sci...358..911A,2019PhRvD.100l3015D}, could also affect primary and secondary lepton spectra in different and very non trivial ways. These complications go beyond the scope of this analysis.

Secondly, we consider uncertainties from the choice of CR projectiles, targets, and cross sections.
Any uncertainty on the CR fluxes and production cross sections directly impact the number of secondary positrons. For CR fluxes, we take demodulated proton and helium CR fluxes measured by AMS-02 \citep{2015PhRvL.114q1103A,2017PhRvL.119y1101A}, and these data typically have uncertainties in the $3-10\%$ range. Also, not accounting for the production from heavier CRs and species heavier than He in the ISM underestimates the secondary positron flux. Following the detailed analysis of \citet{Boudaud:2019efq} carried out for antiprotons, we can estimate these effects to be $\sim 10\%$ and $\sim 3\%$ respectively.

Concerning production cross-section uncertainties, we recall that we use \citet{KamaeEtAl2006} parametrisation. More recent values exist \citep{KachelriessEtAl2019}---they are calibrated on more recent collider data and include incident and target species up to Fe---, but they are only valid for incident nucleus energy greater than 4~GeV. Our analysis is mostly sensitive to the low-energy part, so the latter model is only used to get a rough estimate of the theoretical uncertainties in the production cross sections. The secondary positron flux is 10-20\% larger with \citet{KachelriessEtAl2019} than with \citet{KamaeEtAl2006} values.

Thirdly, we consider uncertainties from energy loss modelling. Positrons suffer different energy losses at high-, intermediate- and low-energies. Above a few tens of GeV, inverse Compton and synchrotron radiation losses have the shortest timescales. Below a few MeV, ionisation and Coulomb losses dominate, and in-between, Bremsstrahlung losses dominate. However, some of the positrons measured below 10~GeV have been produced at higher energy and at a more distant place. Hence, there is no one-to-one correspondence between the hierarchy in energy-loss timescales and impact on the positron flux. This motivates the detailed study of the impact of the various ingredient entering these losses.

Changing the interstellar radiation field (inverse Compton losses) and the magnetic fields (synchrotron losses) according to the values bracketed in \citet{2010A&A...524A..51D} have a $\sim10\%$ impact at~1 GeV.
Coulomb losses on free electrons only dominate at very low energy, and we checked that the uncertainty on $n_e=0.033\pm0.002$~cm$^{-3}$ \citep{1992AJ....104.1465N}---see \citet{2017ApJ...835...29Y} for an updated model---has a negligible impact on the positron flux (sub-percent level) at 1~GeV.

Intermediate energies are dominated by Bremsstrahlung losses on the ISM gas. The same gas density is responsible for the production of secondary positrons. There is a further complication as this very gas density also directly impacts the determination of the transport coefficients, and this is discussed in the next paragraph.
The gas density uncertainty is difficult to assess, and it can be probed for instance via the $\gamma$-ray emissivity \citep[e.g.][]{2011A&A...531A..37D}, especially in the light of Fermi-LAT data (\citealt{2012ApJ...750....3A}, \citealt{2015ApJ...806..240C}, and \citealt{2016ApJS..223...26A}). The impact of the choice of different density maps was recently investigated in \citet{2018ApJ...856...45J}: between 2D and 3D gas models, variations by a factor two on column density were found---this factor mostly comes from using an outdated value of the Sun's position in one gas component of the 2D model, and it certainly overestimates the uncertainty on the surface gas  density ($\Sigma_{\rm ISM}$). More realistically, the $H_{\rm I}$ spin temperature is already responsible for a $10\%$ uncertainty in the gas density \citep{2018ApJ...856...45J}, and other sources of uncertainties come from the still debated $X_{\rm CO}$ conversion value \citep{2017A&A...601A..78R} and the dark gas distribution \citep{2005Sci...307.1292G}. For definiteness, we take a benchmark uncertainty of 50\% on $\Sigma_{\rm ISM}$ in the following.
If we consider together the impact on the production and Bremsstrahlung, there should be no net effect in the regime where Bremsstrahlung losses dominate: The gas density cancels out from the integral calculation, as it appears both in the numerator (production) and in the denominator (propagation)---we stress however, that if primary electrons or positrons are considered (no production), their flux now scales with the inverse of the gas density, as found in \citet{Cirelli:2013mqa}. We checked indeed that there is no impact of the gas density for secondary positrons at 1~GeV. However, at higher energies, in a regime where other energy losses dominate, we also find, as expected, a direct scaling with the gas density.

Fourthly, we consider uncertainties from transport coefficient calibration. The positron flux depends on the transport parameters, calibrated on secondary-to-primary ratios. For instance, assuming a know $\Sigma_{\rm ISM}$, the parameter $K_0$ is determined with a $\sim\pm 12\%$ uncertainty \citep{WeinrichEtAl2020}. This leads to a halved uncertainty on the positron flux at very low energy ($\propto 1/\sqrt{K_0}$), but it fully propagates at 1~GeV. However, as discussed in \citet{2010A&A...516A..67M}, any uncertainty on $\Sigma_{\rm ISM}$ directly translates on the transport parameters $K_0$, $V_{\rm c}$, and $V_{\rm A}$, a behaviour also observed in \citet{2018ApJ...856...45J}. A $\pm50\%$ change on $\Sigma_{\rm ISM}$ would thus change $K_0$ (and $V_{\rm A}$) accordingly, which, combined with the impact on the positron production and Bremsstrahlung, leads to an overall $\pm26\%$ uncertainty on secondary positrons at 1~GeV, this number varying with the energy (see above).

\begin{table}[t]
  \centering
  \caption{Error budget on the calculation of secondary positrons. The first column list the quantities varied in the calculation, the second column provide the typical uncertainties on this ingredient, and the last three columns show the corresponding uncertainty on the calculated IS secondary flux of positrons---the gas surface density impacts the calculation in different places, and its impact is broken down below (see text for details). A `+' sign below (instead of `$\pm$') means that our calculation is conservative, that is, the secondary flux would be larger if we were to account for these specific ingredients. The exact numbers slightly depend on the configuration used (\BIG{}, \SLIM{}, or \QUAINT{}) and we report below values from \QUAINT{}.}
  \label{tab:posit_err}
\begin{tabular}{lcc}
  \hline \hline \\[-1em]
  Ingredient     &   Error on   & $\Delta\Phi^{\rm IS}_{e^+}/\Phi^{\rm IS}_{e^+}$ [\%]\\[0pt]
             &       ingredient [\%]        &  at $(10^{-2},~1,~10^2)$~GeV$^\star$    \\
  \hline
            \multicolumn{3}{l}{\bf CR and gas composition}\\
  CR H and He& $\pm10\%$     &  $\pm10\%$    \\
  +~CRs ($Z>2$)&        -      &  $+10\%$      \\
  +~ISM ($Z>2$)&        -      &  $\pm3\%$     \\[5pt]
           \multicolumn{3}{l}{\bf Energy losses}\\
  ISRF& $\#1\rightarrow\#2^\dagger$& $(+0.2\%,\,-2.7\%,\,-4.1\%)$\! \\
  B          & $\pm~1\mu$G & $(\pm0.7\%,\,\,\mp9.5\%,\,\,\mp12\%)$ \\
  $n_e$      & $\pm10\%$     & \,$(\mp0.2\%,\,\,\mp0.7\%,\;\!<\!0.1\%)$ \\[5pt]
           \multicolumn{3}{l}{\bf Transport calibration \& positron production}\\
  $K_0\,\&\,V_{\rm A}$ & $\pm12\%$       & $(\mp7\%,\;\,\,\pm12\%,\;\,\,\mp5\%)$\\[1pt]
  $(d\sigma/dE)^{\rm prod}$ & $+20\%$  & $+20\%$ \\[5pt]
           \multicolumn{3}{l}{\bf Surface density ($\mathbf{\Sigma_{\rm ISM}}$)}\\
$\Sigma_{\rm ISM}^{\rm PB\,\equiv\, Prod.\&Brem.}$ & $\pm50\%$ & $(\pm13\%,\;\,\pm0.7\%,\,\pm49\%)$  \\[1pt]
$\Sigma_{\rm ISM}^{\rm All~(PB\,\&\,K_0\,\&\,V_{\rm A})}$ & $\pm50\%$ & $(\mp0.7\%,\,\pm26\%,\,\pm21\%)$ \\[5pt]
           \multicolumn{2}{l}{\bf Solar modulation (TOA fluxes)} &  $\mathbf \Delta\Phi^{\rm TOA}_{e^+}/\Phi^{\rm TOA}_{e^+}$ [\%]\\
$\phi$       & $\pm15\%$ & (n/a$^\ddagger$,\;\;\;\;\,$\mp50\%,\;\,\,<1\%)$ \\
  \hline\\[-1em]
\end{tabular}
\footnotesize{\\
$^\star$ For energy-dependent effects, we report 3 values in the table, otherwise a single value is provided.\\
$^\dagger$ Two parametrisations taken from \citet{2010A&A...524A..51D}.\\
$^\ddagger$ At very low energy, the TOA flux is strongly suppressed and its variation is not very meaningful to report.}
\end{table}

Fifthly, we consider uncertainties from Solar modulation. Solar modulating the calculated positron flux also brings uncertainties. Above 100~GeV, CR fluxes are mostly unmodified, and TOA CRs below 1~GeV mostly come from CRs at $\sim$~GeV. As a result, the error budget at 1~GeV applies to lower energies as well for TOA positrons. 

Modulation levels for most data in this analysis were taken from \citet{2017AdSpR..60..833G}, that is, from averages---over the appropriate CR data taking periods---of time series based on the analysis of neutron monitor data. The fact that post-fit values in the LiBeB analyses (see the companion paper, \citealt{WeinrichEtAl2020}) were found to be consistent with the above assumed values further support this choice. In practice, however, \citet{2017AdSpR..60..833G} time series do not extend after 2017. To derive the positron modulation level $\phi_{\rm FF}=650$~MV, we relied on Oulu times series\footnote{\url{http://cosmicrays.oulu.fi/phi}}, derived from \citet{2005JGRA..11012108U}, and rescaled according to $\phi_{\rm Ghelfi17}(t)\approx \phi_{\rm Usoskin05}(t)+ 100~{\rm MV}$, as found in \citet{2017AdSpR..60..833G}.
The overall uncertainties on reconstructed modulation levels from neutron monitors are $\pm 100$~MV \citep{2017AdSpR..60..833G}. Those obtained from directly fitting TOA (pre-AMS) H and He data are in the $\pm30$~MV range \citep{2016A&A...591A..94G}. We choose a very conservative approach below, and for 1~GeV secondary positrons, the $\pm 100$~MV ($\pm 15\%$) uncertainty translates into a $\mp 50\%$ uncertainty. This makes modulation the dominant source of uncertainty for the positron flux calculation.

To conclude this section, we summarise our results on the uncertainties. Although the above analysis does not reach the level of refinement developed for CR nuclei analyses \citep{2019A&A...627A.158D}, we now have a quantitative grasp on the uncertainties on the secondary positron calculations. They are gathered in Table~\ref{tab:posit_err}, where we also provide a finer view of these uncertainties at three energies (10~MeV, 1~GeV, and 100~GeV). At 1~GeV, which is the energy that is relevant for the analysis below, uncertainties from Solar modulation are the dominant effect, followed by those on production and $\Sigma_{\rm ISM}$. Regarding production, we nevertheless stress that the assumptions we make are conservative to derive lower limits on $L$; they underestimate the positron flux by $\sim20-30\%$ (10-20\% from cross sections and 10\% from unaccounted for heavy CR projectiles).

\subsubsection{Constraints on $L$}
To check the compatibility with the limits set by AMS-02 Be/B data, we carried out calculations for several $L$ values. The latter are taken inside their allowed range (upper-half of Table~\ref{tab:Geo_Lr}). To ensure a consistent calculation, the transport parameters are rescaled with $L$ according to Eq.~(\ref{app:rescaling:fit_func}) and values in Table~\ref{tab:scaling} (see App.~\ref{app:rescaling}).

We set conservative limits on $L$ by excluding propagation setups for which the prediction of the secondary positron flux exceeds the AMS-02 measurements \citep{2019PhRvL.122d1102A} by more than $3\sigma$ in a single bin. In practice, for all our configurations, the lowest-energy AMS-02 data point is the one setting constraints.

\begin{figure*}[t]
  \includegraphics[width=0.33\textwidth]{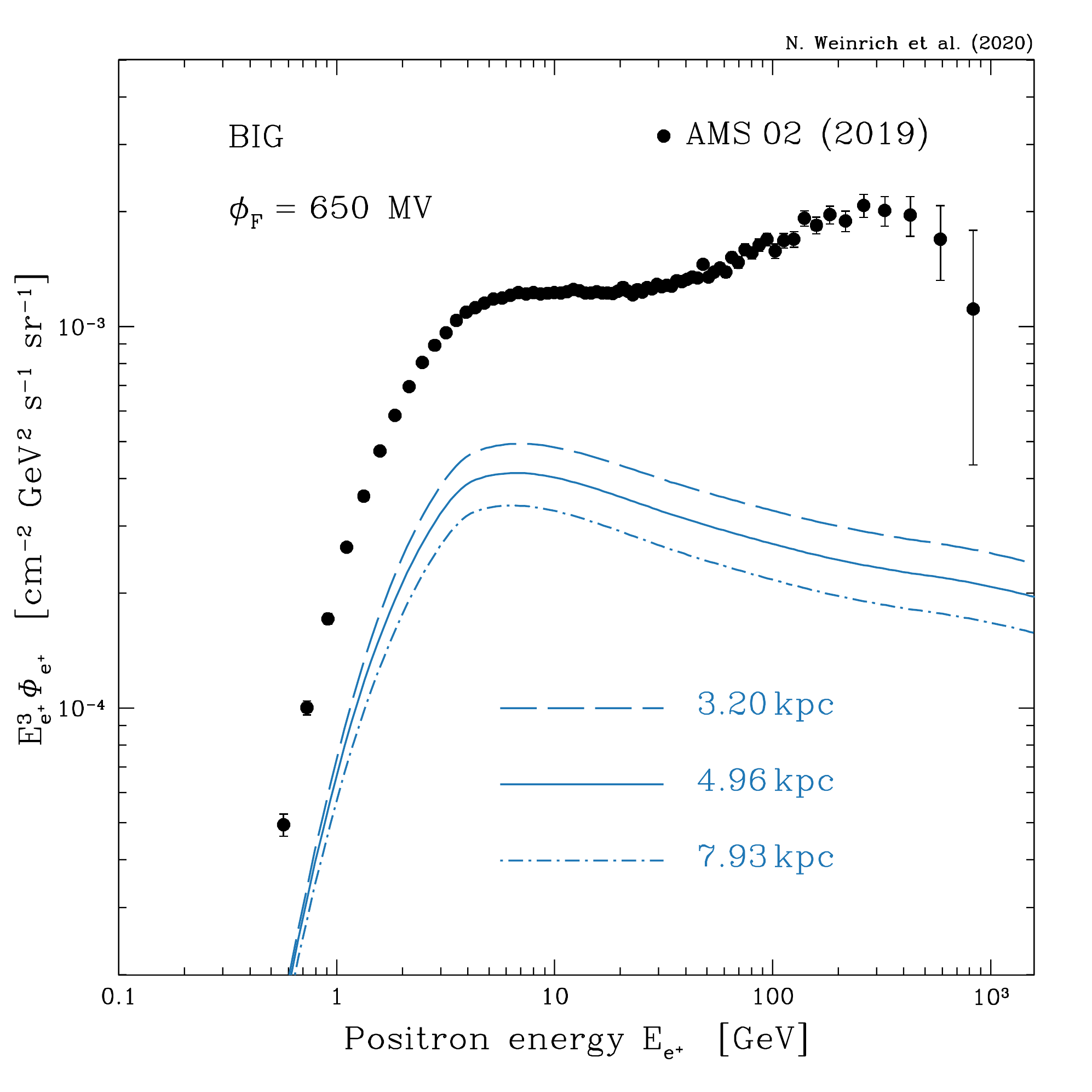}
  \includegraphics[width=0.33\textwidth]{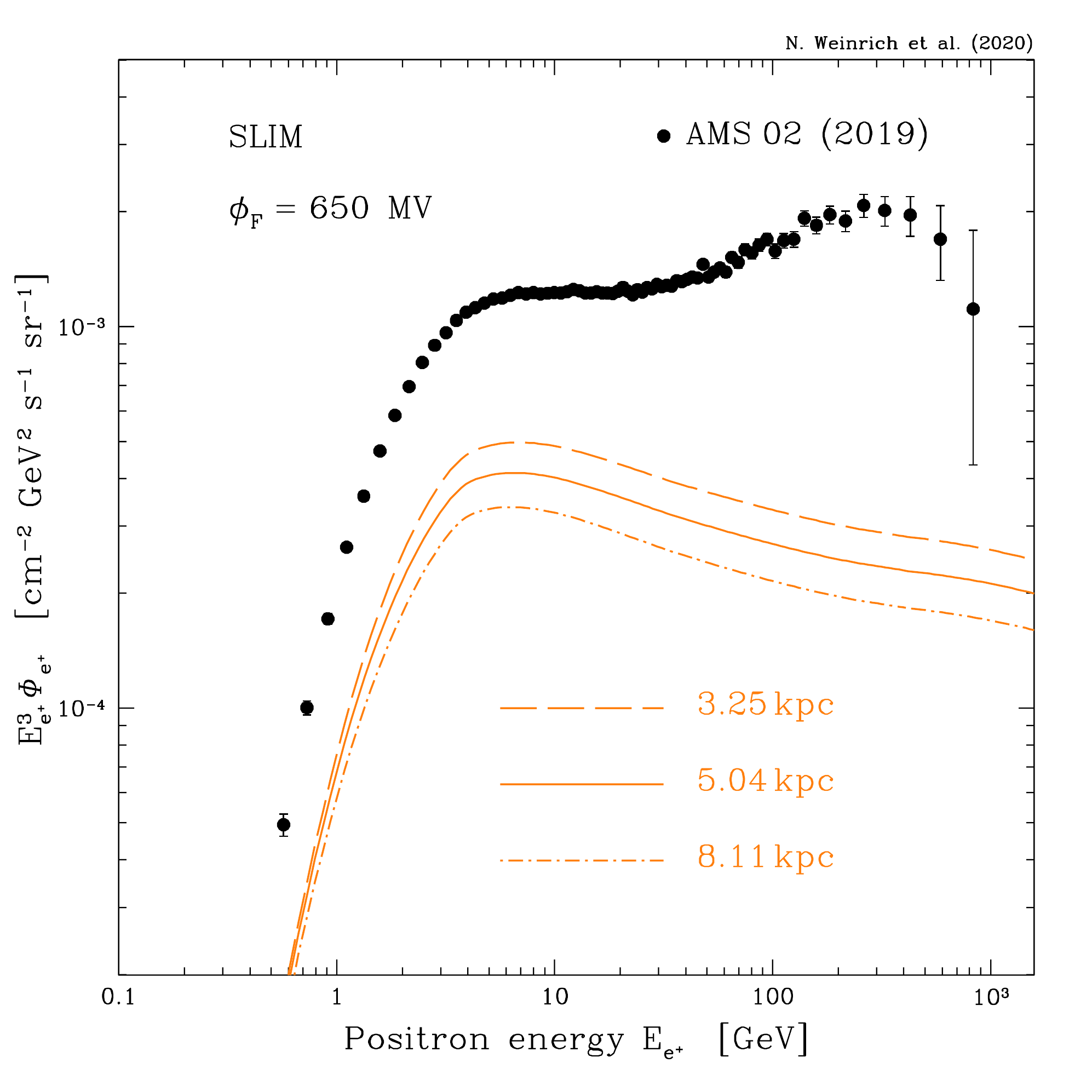}
  \includegraphics[width=0.33\textwidth]{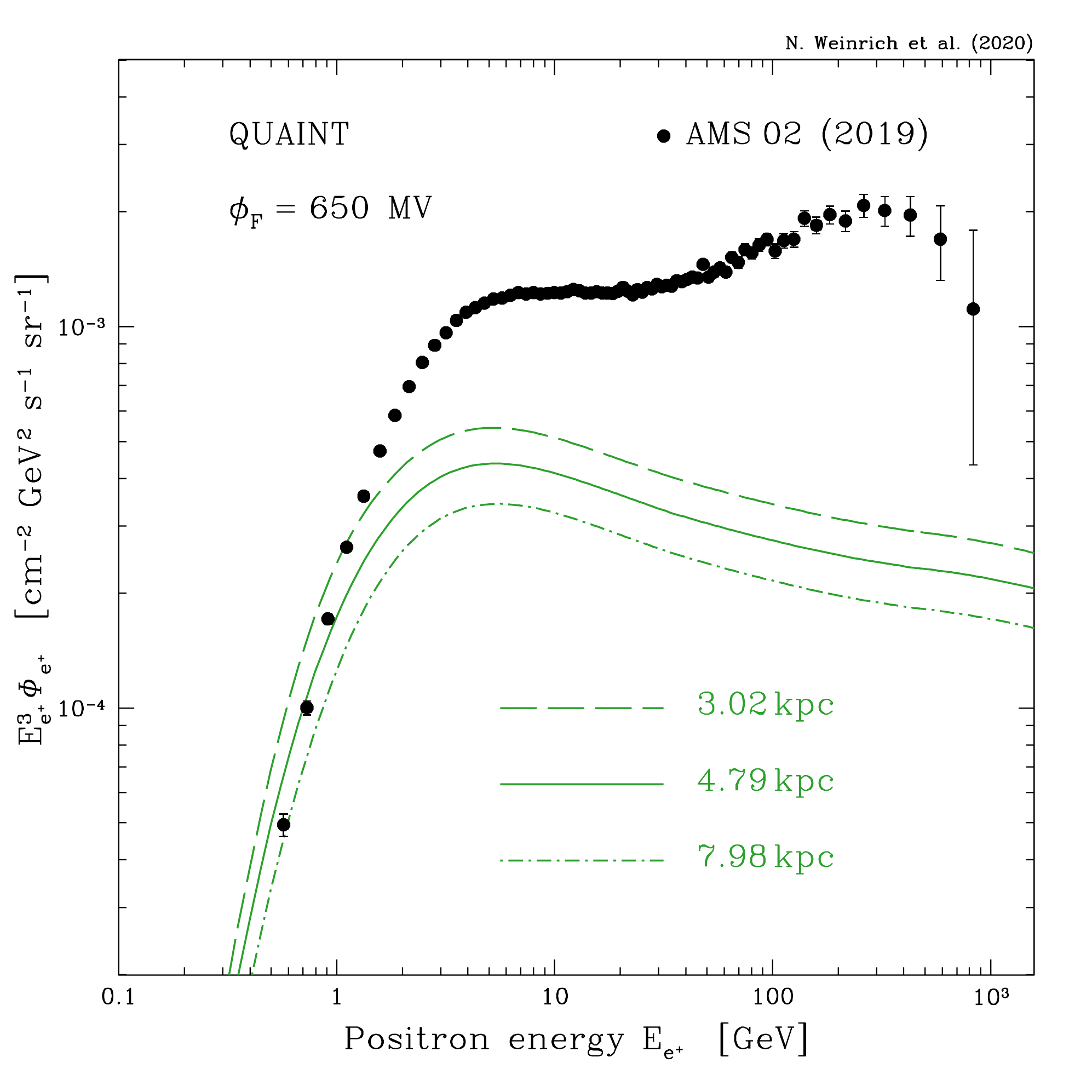}
  \caption{Secondary positron flux predictions using the best-fit transport parameters of the \BIG{} (left panel), \SLIM{} (middle panel), and \QUAINT{} (right panel) propagation models, as inferred from the AMS-02 combined Li/C, Be/B, and B/C analysis. The three lines correspond to TOA fluxes, modulated at 650~MV, for the best-fit (solid line) and $1\sigma$ upper (dash-dotted) and lower (dashed) limits on $L$ (see Table~\ref{tab:Geo_Lr})---AMS-02 data are from \citet{2019PhRvL.122d1102A}.}
  \label{fig:positron_flux}
\end{figure*}

In a first step, we check the consistency of the positron flux with the Be/B constraint. In \citefig{fig:positron_flux}, we report our predictions for the secondary positron fluxes (rescaled by  $E^3$) for the \BIG, \SLIM, and \QUAINT\ propagation models (left, middle, and right panel, respectively) along with the AMS-02 data \citep{2019PhRvL.122d1102A}. These predictions are derived using the best-fit parameters inferred from the combined Li/C, Be/C, and B/C analysis performed in \citet{WeinrichEtAl2020}, already extensively discussed throughout the paper. For illustration, we have taken the best-fit values obtained for $L$ from the AMS-02 Be/B data only (see Table~\ref{tab:Geo_Lr}). In each panel, the solid curve corresponds to the best-fit value for $L$, bracketed by its $1\sigma$ statistical uncertainty (dashed curve for the lower deviation, and dot-dashed curve for the upper deviation).

From \citefig{fig:positron_flux}, we see that the positron constraint on $L$ is only relevant for the \QUAINT\ propagation model. The latter has an effective Alfv\`en speed $V_{\rm A}\sim 40$~km/s (see Table~\ref{tab:scaling}) which sets the amplitude of reacceleration\footnote{We stress that pre-AMS-02 fit to B/C data (e.g. \citealt{2001ApJ...555..585M}) had larger $V_{\rm A}$. It was found that strong reacceleration gave rise to a prominent bump around 1~GeV in the predicted positron flux, which then easily overshoots the data, especially below a few GeV (\citealt{DelahayeEtAl2009}, \citealt{2014PhRvD..90h1301L}, and \citealt{BoudaudEtAl2017a}). However, with milder reacceleration here, we see in the right panel that the flux predictions associated with \QUAINT\ are not in that strong excess with respect to the data, and do not feature any significant bump.}.
For the \BIG\ and \SLIM\ models (left and middle panels), both devoid of reacceleration, the nominal predictions undershoot the data. The low-energy behaviour in these models is furthermore exacerbated by the low-rigidity break in the diffusion coefficient around $3-4$~GV \citep{WeinrichEtAl2020}: it leads to a lower secondary positron flux at low rigidities and hence a decreased constraining power on $L$.
Given the uncertainties on the calculation, only \QUAINT{} can overshoot the data, whereas \BIG{} and \SLIM{} cannot for $L$ values constrained by Be/B.

\begin{figure}[!t]
  \includegraphics[width=\columnwidth]{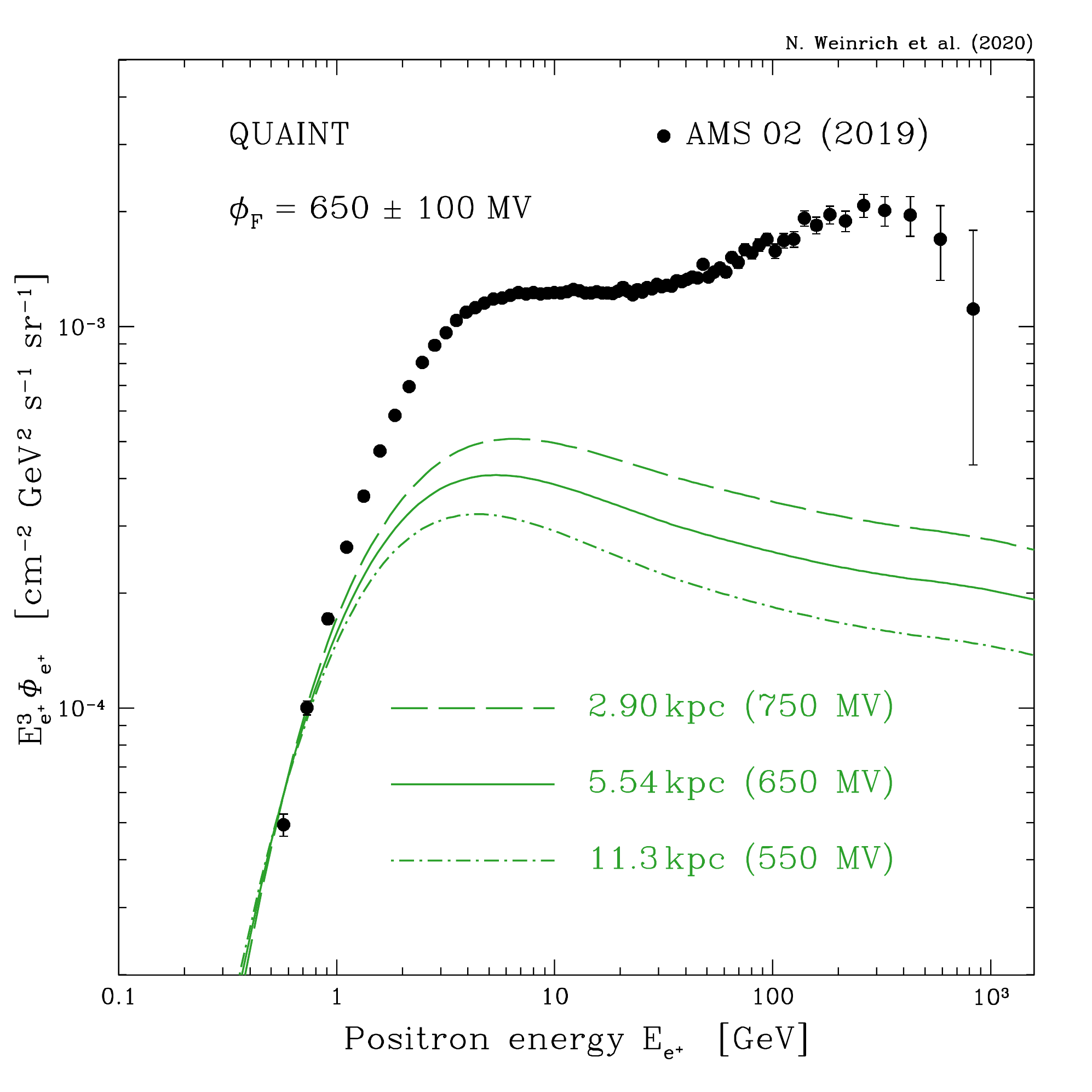}
  \caption{Secondary positron flux predictions using the best-fit transport parameters of the \QUAINT\ propagation model (also inferred from the AMS-02 combined Li/C, Be/B, and B/C analysis). We show the predictions associated with the $3\sigma$ limits on $L$ assuming different solar modulation strengths.}
  \label{fig:positron_limits}
\end{figure}

In a second step, we derive the upper-limit on $L$ in the model \QUAINT{}. Our nominal $3\sigma$ constraint for \QUAINT{} is $L\geq 5.54$~kpc. However, this number is very sensitive to the model uncertainties. For illustration, we show in Fig.~\ref{fig:positron_limits} the impact of Solar modulation uncertainties, varying the Fisk potential by a generous $\pm 100$~MV \citep{2017AdSpR..60..833G}. The upper value of the modulation, which minimises the modulated positron flux, should be viewed as conservative, weakening the limit to $L\geq 2.9$~kpc. The lower value corresponds to a more aggressive constraint (for a lower solar activity) $L\geq 11.3$~kpc. These limits do not account for sub-leading uncertainties discussed above.
Without a better handle on the uncertainties, it is difficult to firmly conclude. Overall, we note that the positron constraint on $L$ for the \QUAINT\ model is in agreement with those derived from the Be/B and \tentonineBe{} analysis, while being independent from the latter.

We have only discussed the case of overshooting, which provides a clear situation for excluding part of the parameter space. Given that in the \BIG\ and \SLIM\ models the nominal predictions undershoot the data, one may wonder if their viability is questioned by the positron data.
We remark that the various sources of uncertainties (modulation, production, surface gas density), estimated at the $\lesssim50\%$ level at most, seem unable to fully account for the factor of $\sim 2$ mismatch.
However, primary positrons, already necessary to explain the high-energy positron fraction in pre-AMS-02 studies, might make up a significant fraction of the positron budget also at low energy---for example, the absence of primary positrons at low energy was raised as an important issue for astrophysical sources like pulsar wind nebulae in \citet{2011ASSP...21..624B}.
Therefore, discussing further the consistency of any propagation model (not only \BIG{} and \SLIM{}) with the positron data should rely on analyses also including primary positrons. This goes beyond the scope of this paper.

\section{Conclusions}
\label{sec:conclusions}

In the context of recent high-precision AMS-02 data, we have revisited the constraints set on the halo size of the Galaxy from radioactive species and positron fluxes.

Using AMS-02 Be/B data we find $L=5^{+3}_{-2}$~kpc at $1\sigma$, in agreement with \citet{2020ApJ...889..167B} (but with a larger error) and less than one sigma away from the 
 results of \citet{2020PhRvD.101b3013E}. Our result holds for several transport configurations \citep{2019A&A...627A.158D,WeinrichEtAl2020}, namely \BIG{} (reacceleration and low-rigidity break), \SLIM{} (pure diffusion with low-rigidity break), and \QUAINT{} (reacceleration and diffusion upturn in the non-relativistic regime). The constraints are tighter (factor $\sim 2$ reduction in errors) and move to a lower value of $L$ (by 0.3-0.7 kpc) when low-energy \tentonineBe{} data are considered, but this tightening may be related to the fact that ISOMAX data \citep{2004ApJ...611..892H} prefer a smaller halo size $L\approx 3$~kpc and might be indicative of a tension in the data. With the recent release of \het{} and \hef{} data \citep{2019PhRvL.123r1102A}, AMS-02 demonstrated its capabilities for measuring isotopic fluxes. Separating Be isotopes will certainly be even more challenging, but AMS-02 will provide a unique picture in the few GeV/n regime, where only ISOMAX and preliminary PAMELA data \citep{2019BRASP..83..967B} are available for now, and shed some light on the mutual consistency of these datasets. The balloon-borne HELIX, should also provide a complementary view. The instrument is scheduled for a long-duration balloon flight in 2020/21, and is expected to achieve a 10\% statistical error on \tentonineBe{} in the 0.1-10 GeV/n range \citep{2019ICRC...36..121P}.

We have also performed a detailed analysis of the modelling uncertainties for both the Be/B and \tentonineBe{} ratios to understand whether better data could help improving the estimation on $L$ in the future. Whereas Be/B is maximally sensitive to $L$ at a few tens of GV, \tentonineBe{} is sensitive to $L$ over a much larger range (from 100~MeV/n to tens of GeV/n). In terms of possible improvements, Be/B data are already limited by systematics, whereas this is not yet the case for \tentonineBe{}. For these reasons, \tentonineBe{} seems to be the best target for future experiments. However, in both cases, production cross-sections uncertainties dominate the modelling error budget and are already at the level of data uncertainties. Future improvements on $L$ will thus not be possible without improving nuclear data. An alternative strategy to mitigate these uncertainties would be to combine data from different CR clocks (e.g. Al/Mg and Cl/Ar).

In a broader context of multi-wavelength and multi-messenger observations, we also discussed the constraints set by synchrotron radio \citep[e.g.][]{2013JCAP...03..036D} and diffuse $\gamma$-ray emissions \citep[e.g.][]{2015ApJ...807..161T} in the Milky-way (and in some cases in other galaxies).
 The various observations lead to lower or upper limits, defining a broad range $L\in[2,10]$~kpc compatible with results from radioactive CR nuclei.  We also updated the constraints set by positrons \citep{2014PhRvD..90h1301L}. Since \BIG{} and \SLIM{} configurations undershoot the data, the constraints are only significant for the \QUAINT{} model, leading to $L\geq 2.9-11.3$~kpc depending on the solar modulation, with a nominal central value $L\geq 5.5$~kpc; these numbers could shift upwards or downwards depending on uncertainties on the production cross section and the gas surface density. Within the errors, these constraints are also consistent with the ones derived in the main analysis.

While our conclusions appear rather robust within the framework of this analysis, one should keep in mind that these bounds might be altered in presence of inhomogeneities in the local gas density (for radioactive species) and on the diffusion coefficient (for both radioactive species and positrons). These extensions represent interesting and motivated subjects for future studies.

\begin{acknowledgements}
We thank A.~Marcowith for discussions. This work has been supported by Univ.~de Savoie, appel à projets: {\em Diffusion from Galactic High-Energy Sources to the Earth (DIGHESE)}, by the Programme National des Hautes Energies of CNRS/INSU with INP and IN2P3, co-funded by CEA and CNES, and by Villum Fonden under project no.~18994. This work is partly supported by the ANR project ANR-18-CE31-0006, the national CNRS-INSU programmes PNHE and PNCG, and European Union's Horizon 2020 research and innovation programme under the Marie Sk\l{}odowska-Curie grant agreements N$^\circ$ 690575 and N$^\circ$ 674896.
\end{acknowledgements}

\appendix

\section{Scaling relations with $L$}
\label{app:rescaling}

In Sect.~\ref{subsec:expected}, we study the sensitivity to $L$ of CR data combinations involving a radioactive species. To do so, the transport parameters obtained from the study of B/C must be known for any $L$. In Sect.~\ref{sec:L_posit},
we test whether secondary positrons overshoot the data given for various $L$, given the constraints set from secondary-to-primary ratios and radioactive species. This is a similar but slightly different question than the previous one.

\begin{figure}[!t]
  \label{app:rescaling:fit_results}
  \includegraphics[width=\columnwidth]{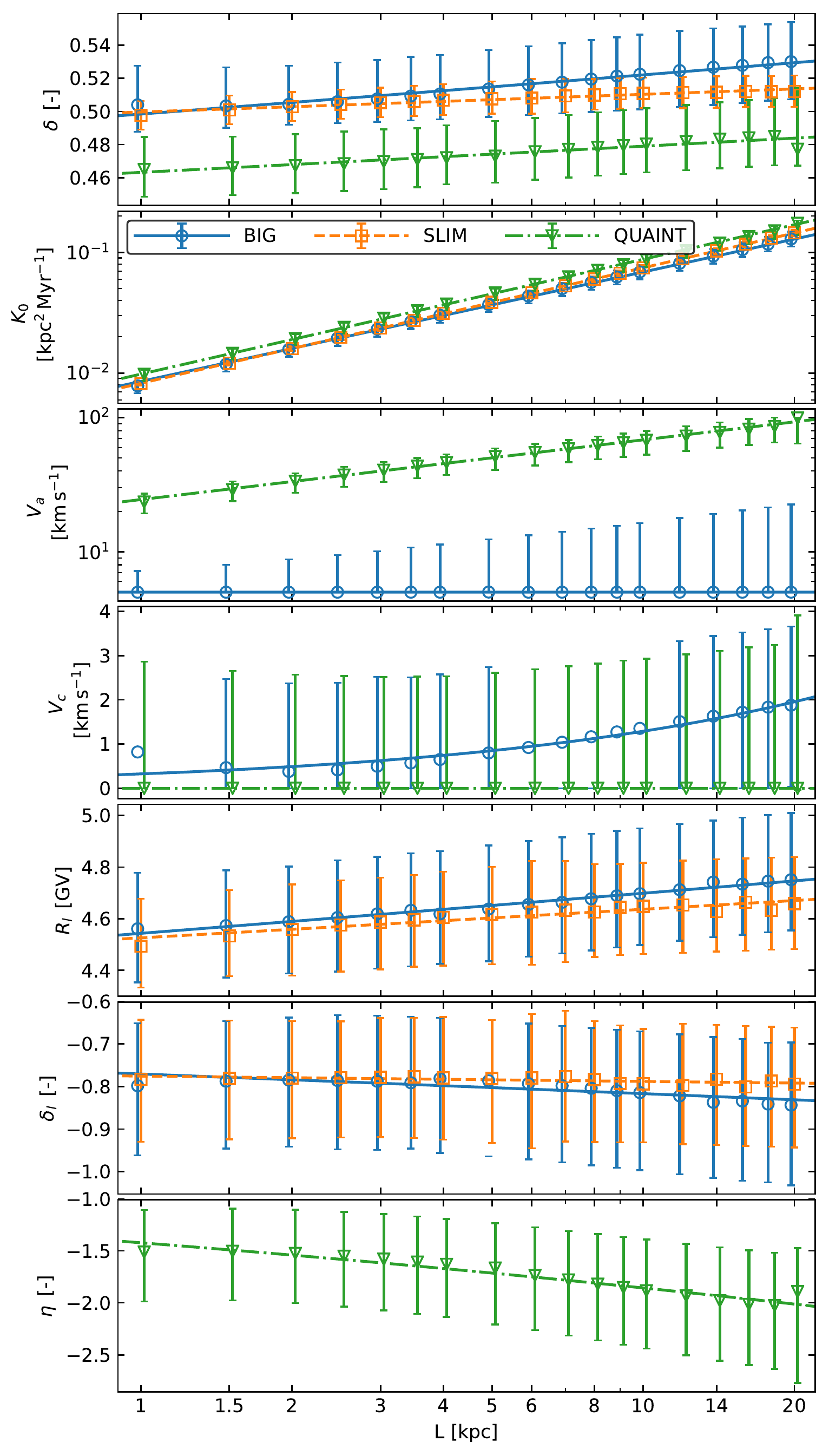}
  \caption{Transport parameters dependence on the halo size $L$. Colour-coded symbols correspond to the best-fit values on the combined analysis of Li/C, Be/B, and B/C data for different models (\BIG{}, \SLIM{}, and \QUAINT{}). The lines show, for each model and transport parameter, power-law fits to these data whose values are reported in the right-hand side of Table~\ref{tab:scaling}.}
\end{figure}
For each model (\BIG{}, \SLIM{}, and \QUAINT{}), we repeat the transport parameter fitting procedure described in \citet{WeinrichEtAl2020} {\em at fixed $L$} for several values of $L$, for either the AMS-02 B/C ratio only, or the combined AMS-02 Li/C, Be/B, and B/C data. We show for the latter the resulting best-fit transport parameters (symbols) in Fig.~\ref{app:rescaling:fit_results}. We find that a simple power law is enough to capture the dependence with $L$ (lines in Fig.~\ref{app:rescaling:fit_results}),
\begin{equation}
\label{app:rescaling:fit_func}
  \displaystyle
  \mathrm{Param}(L) = A \cdot \left( \frac{L}{5\,\mathrm{kpc}} \right)^{B}\,,
\end{equation}
and we report the best-fit $A$ and $B$ values found for each transport parameter ($\delta$, ${K}_0$, etc.) in Table~\ref{tab:scaling} for the two different cases.
\begin{table*}[t]
\center
  \caption{Scaling coefficients for the transport parameters, see Eq.~(\ref{app:rescaling:fit_func}), with $B$ the power-law slope and $A$ the transport parameter value for $L=5$~kpc. Two analyses are reported, based on the fit of AMS-02 B/C data only (left-hand side), or for the combined fit of AMS-02 Li/C, Be/B, and B/C data (right-hand side).}
  \label{tab:scaling}
\begin{tabular}{lccrrrcrrr}\hline \hline \\[-1em]
	Parameter & Coef.& \quad & \multicolumn{3}{c}{Fit B/C}& \quad\quad\quad & \multicolumn{3}{c}{Fit Li/C, Be/B, and B/C}\\
            &      &       & \quad\BIG{}\quad & \quad\SLIM{}\quad & \QUAINT{} & & \quad\BIG{}\quad & \quad\SLIM{}\quad & \QUAINT{}\\
	\hline \\[-1em]
	\multirow{2}{*}{\shortstack{ $\delta \,$ [-] } } 
	& A & &  0.488 &  0.511 &  0.458  & & 0.515 & 0.507 &  0.474   \\ 
	& B & & -0.013 & -0.011 & -0.013  & & 0.020 & 0.009 &  0.015   \\[4pt]
	\multirow{2}{*}{\shortstack{ $K_0\;$  $ \mathrm{[kpc^2\, Myr^{-1}]}$ } } 
	& A & &  0.048 &  0.043 &  0.056  & & 0.037 & 0.038 &  0.045   \\ 
	& B & &  1.043 &  1.034 &  1.040  & & 0.907 & 0.957 &  0.952   \\[4pt]
	\multirow{2}{*}{\shortstack{ $V_{\rm A}\,$ $\mathrm{[km\, s^{-1}]}$ } } 
	& A & &  42.94 & n/a    &  67.24  & & 5.001 & n/a   &  50.19   \\ 
	& B & &  0.536 & n/a    &  0.520  & & 0.000 & n/a   &  0.445   \\[4pt]
	\multirow{2}{*}{\shortstack{ $V_{\rm c} \,$ $\mathrm{[km\,s^{-1}]}$ } } 
	& A & &  0.010 & n/a    &  0.100  & & 0.851 & n/a   &  0.000   \\ 
	& B & & -0.008 & n/a    &  0.000  & & 0.600 & n/a   &  0.000   \\[4pt]
	\multirow{2}{*}{\shortstack{ $R_l \,$ [GV] } } 
	& A & &  3.605 &  4.393 & n/a     & & 4.651 & 4.603 &  n/a     \\ 
	& B & & -0.000 &  0.008 & n/a     & & 0.015 & 0.010 &  n/a     \\[4pt]
	\multirow{2}{*}{\shortstack{ $\delta_l \,$ [-] } } 
	& A & & -0.525 & -0.696 & n/a     & & -0.803&-0.784 &  n/a     \\ 
	& B & & -0.024 & -0.016 & n/a     & &  0.025& 0.007 &  n/a     \\[4pt]
	\multirow{2}{*}{\shortstack{ $\eta \,$ [-] } } 
	& A & & n/a    & n/a    & -0.140  & & n/a   & n/a   & -1.713   \\ 
	& B & & n/a    & n/a    & -0.208  & & n/a   & n/a   &  0.116   \\ 
	\hline
\end{tabular}
\end{table*}

\subsection{Power-law behaviour of the rescaling}
\begin{figure}[!t]
    \includegraphics[width=\columnwidth]{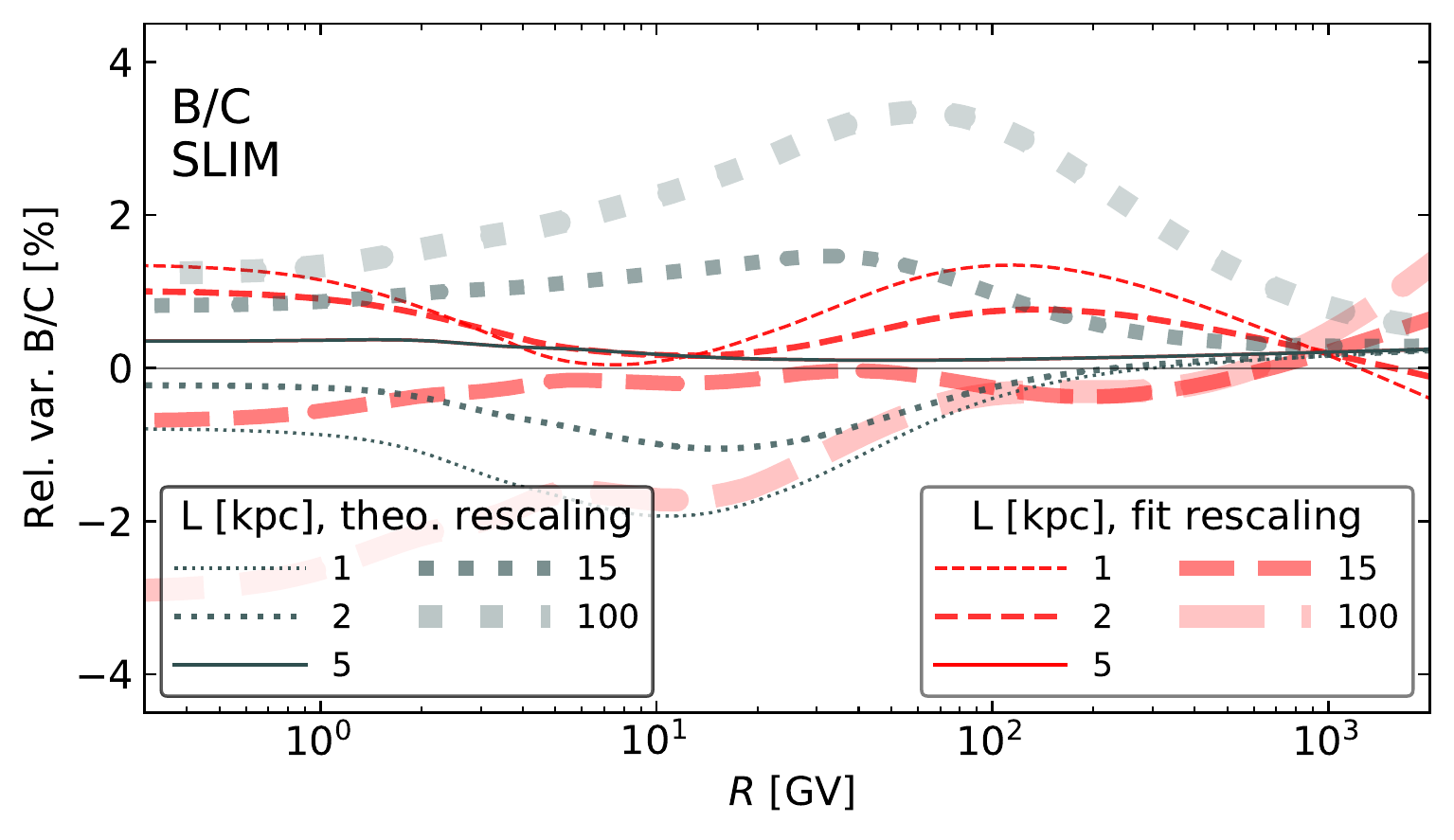}
    \caption{Relative variation of B/C ratio (with respect to a reference taken at $L=5$~kpc), as a function of rigidity. Thin- to thick-dashed lines correspond to calculations for various $L$ using either (i) the `theoretical' rescaling which enforce the same level of production for stable secondary species (grey lines), or (ii) the rescaling based on a refit of B/C data for each $L$ as described in the text (red lines).}
    \label{fig:rescaling_fit_vs_theory}
\end{figure}

A few comments are in order: the power-law indices, $B$, for the B/C analysis are very similar to the ones found in Fig.~5 of \citet{2010A&A...516A..66P}, that is 1.06 for $K_0$ and 0.53 for $V_{\rm A}$ (from the analysis of older B/C data). As argued in \citet{2010A&A...516A..66P}, a secondary-to-primary ratio provides constraints degenerated with $K_0/L$ and $V_{\rm A}^2/L$ ratios. These small differences (1.06 vs 1 and 0.53 vs 0.5) originate in the small fraction of $^{10}$B ($\lesssim 10\%)$ coming from decayed \tenBe{}, which is sensitive to $L$. Figure~\ref{fig:rescaling_fit_vs_theory} illustrates this sensitivity on the B/C ratio. If the theoretical rescaling is applied (grey lines), the curves with varying $L$ are either above or below the reference, converging to the reference at high rigidity as \tenBe{} half-life grows and \tenBe{} behaves as a stable secondary species. At variance, when the above rescaling is applied, the variation with $L$ is absorbed in the fit, and the high-rigidity limit no longer goes to zero. The overall difference is at the few percent level.

\subsection{Difference between B/C and combined analysis results}
A similar scaling remains for the combined analysis (Li/C, Be/B, B/C), but with some differences, that were all highlighted in \citet{WeinrichEtAl2020}. First, the diffusion coefficient normalisation $K_0$ is smaller than in the B/C only case, and the difference is related to partial degeneracies with production cross sections, that are lifted in the combined analysis of elements. The best-fit values are consistent, though slightly different, from those in \citet{WeinrichEtAl2020}, and this is mostly attributed to the fact that we fit here the combination Be/B instead of Be/C. Second, whereas for the B/C case \BIG{} parameters are very similar to \QUAINT{} ones, for the combined analysis the \BIG{} parameters are closer to \SLIM{} ones. Last, the low-energy parameters $\delta_l$ and $\eta$ are also different from the B/C analysis only, and correspond to more marked low-rigidity break of the diffusion coefficient; we find here that the latter parameters have almost no dependence on $L$.

\section{$L$ from several $^{10}$Be/$^9$Be datasets}
\label{app:Lper10Bedataset}

The \tentonineBe{} ratio has been measured by different experiments in different energy ranges. To better grasp the respective weight of the data in the final constraint (see Sect.~\ref{sec:constraintL_other}), we show the different contributions to the fit. We focus on \SLIM{}, but similar results are observed for \BIG{} and \QUAINT{}.

\begin{figure}[!t]
  \includegraphics[width=\columnwidth]{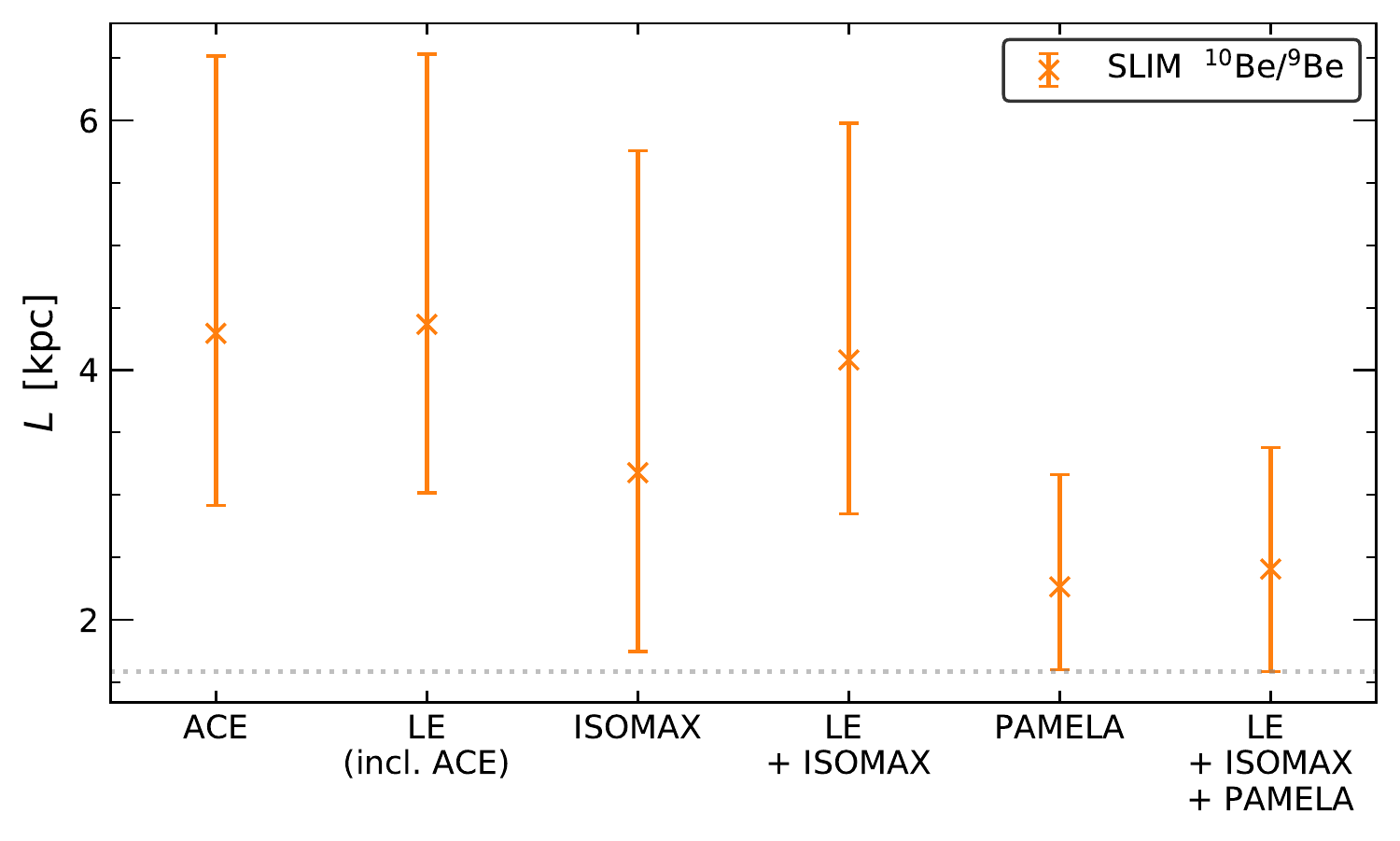}
  \caption{Constraints on $L$ (top panel) for combined fit of AMS-02 Li/C and B/C data, and various datasets for \tentonineBe{}. From left to right: low-energy ACE data, all low-energy data combined (ACE, IMP7\&8, ISEE3, Ulysses-HET, and Voyager), higher-energy GeV/n ISOMAX \citep{2004ApJ...611..892H}, LE and ISOMAX combined (fourth column), preliminary PAMELA data \citep{2019BRASP..83..967B}, and all combined (last column).}
  \label{fig:Be-PAMELA}
\end{figure}
In Fig.~\ref{fig:Be-PAMELA}, we show the combined analysis of Li/C, B/C, and different datasets for \tentonineBe{}. Fits on low-energy data only are completely driven by ACE data (compare first and second column). They lead to larger $L$ values than the higher-energy ISOMAX data (third column). The latter have larger uncertainties than low-energy data (see Fig.~\ref{fig:Be-contours}), so that low-energy data drive the combined constraint (fourth column).

For completeness, we also show the constraints from the preliminary PAMELA data \citep{2019BRASP..83..967B}. They cover a energy range of 0.1 to 2 GeV/n, combining two independent analyses with the TOF and calorimeter, with a $\sim 20\%$ uncertainty. They give the best constraints on $L$ (next-to-last column) with smaller error bars, in line with the expectations discussed of Sect.~\ref{subsec:expected}. Most importantly, they point towards smaller values of the halo size than the other datasets, with $L\sim 2-3$~kpc. If all \tentonineBe{} data are combined, the PAMELA data drive the fit (last column).

AMS-02 data will cover a similar range as PAMELA preliminary data. With probably slightly smaller uncertainties, they are expected to provide similar or slightly better precision on $L$. Because of the tension between PAMELA and lower-energy data, the results from both experiments are crucial to be able to obtain robust results on the central value of the halo size.

\section{Constraints from $^7{\rm Be}/(^9{\rm Be}+^{10}\!{\rm Be})$}
\label{app:7Beratio}

Owing to the difficulty to achieve isotopic separation in CR experiments, data analyses often start with the most favourable configuration. The Be element is made of \sevenBe{}, \nineBe{}, and \tenBe{}. Taking advantage of the $\Delta A=2$ mass separation between $A=7$ and $A=9,10$ isotopes, the $^7{\rm Be}/(^9{\rm Be}+^{10}\!{\rm Be})$ ratio is experimentally a favourable configuration to analyse.

\begin{figure}[!t]
    \includegraphics[width=\columnwidth]{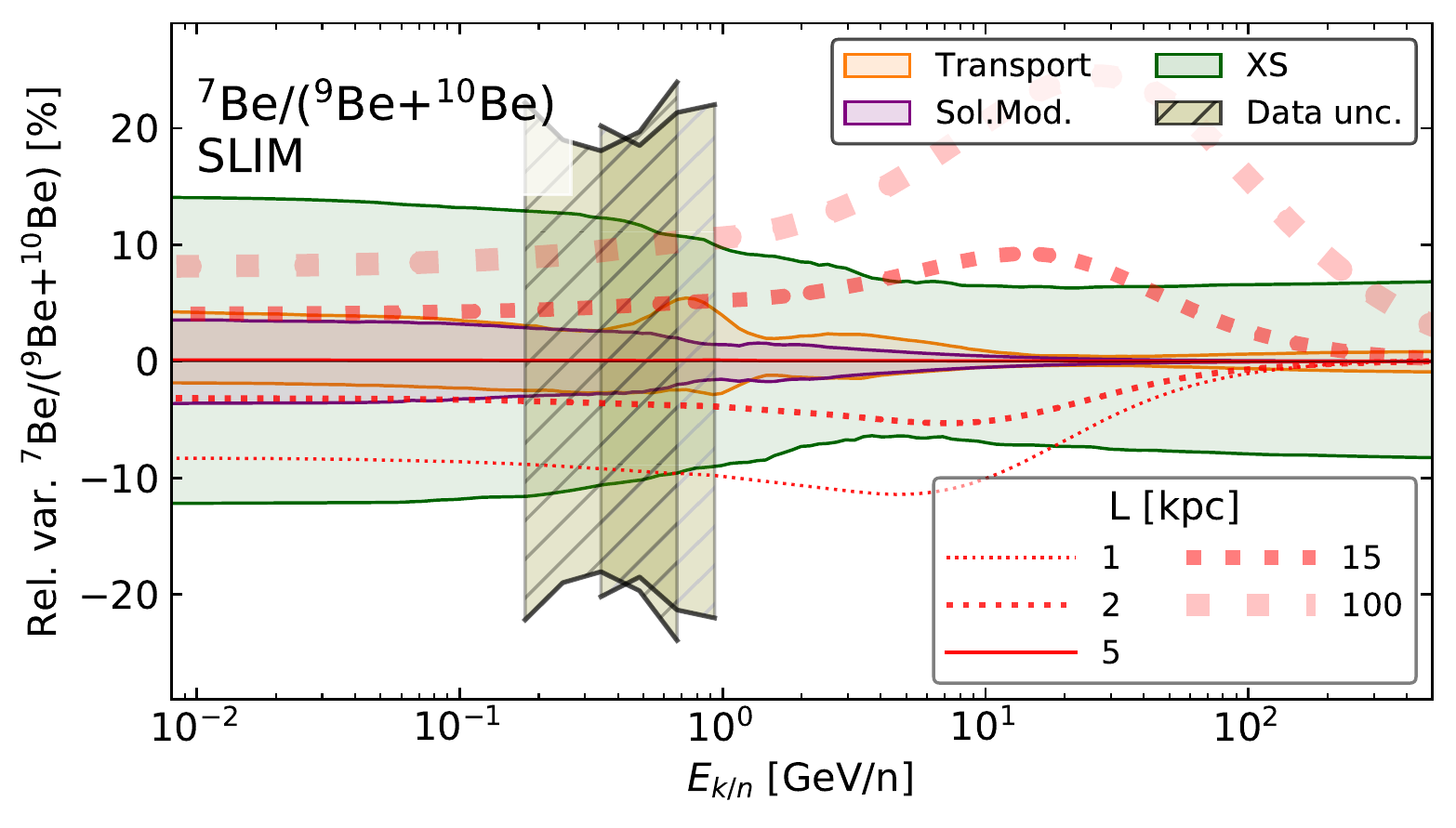}
    \caption{Same as Fig.~\ref{fig:constraints_gilles_plot}, but for the variation of $^7$Be/$(^9$Be$+^{10}$Be) with $L$ (pinkish dashed lines), compared to various model uncertainties (contours) and current data uncertainties (hatched boxes); data from AMS-01 \citep{2011ApJ...736..105A} and PAMELA preliminary analysis \citep{2018ApJ...862..141M}.}
    \label{fig:èBeto9and10Be}
\end{figure}
This ratio has been published by AMS-01 \citep{2011ApJ...736..105A} and also by PAMELA \citep{2018ApJ...862..141M}. Following similar steps as the analysis presented in Sect.~\ref{subsec:expected}, we show in Fig.~\ref{fig:èBeto9and10Be} prospective limits on $L$ that can be set from using the $^7{\rm Be}/(^9{\rm Be}+^{10}\!{\rm Be})$ ratio. Because of the sub-dominant abundance of $^{10}$Be in the denominator, this ratio is as sensitive to $L$ as the Be/B ratio. But whereas AMS-02 achieves a few percent precision on Be/B, current experiment are at $\sim15-20\%$ precision for isotopic ratios.
For this reason, we conclude that $^7{\rm Be}/(^9{\rm Be}+^{10}\!{\rm Be})$ is not a competitive target to fix $L$ with current data or with future experiments.

\bibliographystyle{aa} 
\bibliography{halosize}
\end{document}